\documentclass[aps,pre,twocolumn,superscriptaddress,groupedaddress]{revtex4}  % for review and submission

\usepackage{graphicx}  % needed for figures

\usepackage{url}                  %this package should fix any errors with URLs in refs.
\usepackage{dcolumn}       % needed for some tables
\usepackage{bm}                  % for math
\usepackage{amssymb}       % for math
\usepackage{mathtools}     % for math
\usepackage{pgf}                   %plots from matplotlib in TeX style
\usepackage{tikz}                   %draw graphics
\usepackage{lipsum}

\begin{document}

\title{Echo State Network for two-dimensional turbulent moist Rayleigh-B\'{e}nard convection}
\author{Florian Heyder}
\affiliation{Institut f\"ur Thermo- und Fluiddynamik, Technische Universit\"at Ilmenau, Postfach 100565, D-98684 Ilmenau, Germany}
\author{J\"org Schumacher}
\affiliation{Institut f\"ur Thermo- und Fluiddynamik, Technische Universit\"at Ilmenau, Postfach 100565, D-98684 Ilmenau, Germany}
\affiliation{Tandon School of Engineering, New York University, New York City, NY 11201, USA}
\date{\today}

\begin{abstract}
Recurrent neural networks are machine learning algorithms which are suited well to predict time series. Echo state networks are one specific implementation of such neural networks that can describe the evolution of dynamical systems by supervised machine learning without solving the underlying nonlinear mathematical equations. In this work, we apply an echo state network to approximate the evolution of two-dimensional moist Rayleigh-Bénard convection and the resulting low-order turbulence statistics. We conduct long-term direct numerical simulations in order to obtain training and test data for the algorithm. Both sets are pre-processed by a Proper Orthogonal Decomposition (POD) using the snapshot method to reduce the amount of data. Training data comprise long time series of the first 150 most energetic POD coefficients. The reservoir is subsequently fed by these data and predicts of future flow states. The predictions are thoroughly validated by original simulations. Our results show good agreement of the low-order statistics. This incorporates also derived statistical moments such as the cloud cover close to the top of the convection layer and the flux of liquid water across the domain. We conclude that our model is capable of learning complex dynamics which is introduced here by the tight interaction of turbulence with the nonlinear thermodynamics of phase changes between vapor and liquid water. Our work opens new ways for the {\em dynamic} parametrization of subgrid-scale transport in larger-scale circulation models. 
\end{abstract}

\maketitle

\section{Introduction}
\label{sec:intro}
\noindent
Machine learning (ML) algorithms have changed our way to model and analyse turbulent flows including thermally driven convection flows \cite{BrennerPRF2019,BruntonARFM2020,pandey_perspective2020}. The applications reach from subgrid-scale stress models \cite{Ling2016,DuraisamyARFM2019}, via the detection of large-scale patterns in mesoscale convection~\cite{fonda2019} to ML-based parametrizations of turbulent transport and clouds in climate and global circulation models ~\cite{brenowitz2018, ogorman2018,gentine2018}. The implementations of such algorithms help to process and reduce increasing amounts of data coming from numerical simulations and laboratory experiments \cite{Cierpka2020} by detecting patterns and statistical correlations \cite{Goodfellow-et-al-2016}. Moreover, the computational cost that is in line with a direct numerical simulation (DNS) of the underlying highly nonlinear Navier-Stokes equations can often be reduced significantly by running a neural network instead that generates synthetic fields with the same low-order statistics as in a full simulation \cite{Schneider2017,Mohan2020}. 

Turbulent convection, as all other turbulent flows, is inherently highly chaotic so that specific flow states in the future are hard to predict after exceeding a short horizon. The additional possibility of the fluid to change its thermodynamic phase, as it is the case in moist turbulent convection \cite{StevensMoist2005,MelladoARFM2017}, adds further nonlinearities and feedbacks to the turbulence dynamics. Learning low-order statistics of such a turbulent flow provides a challenge to an ML algorithm. For such a task, an internal memory is required since statistical correlations decay in a finite time. This necessitates the implementation of a short-term memory or internal cyclic feedbacks in the network architecture. That is why a particular class of supervised machine learning algorithms -- recurrent neural networks (RNNs) -- will be in the focus of the present work \cite{SchmidhuberHochreiter1997,LukoseviciusJaeger2009,TanakaNeuralNetworks2019}. 

In this paper we apply a specific implementation of an RNN, the {\em echo state network} (ESN) \cite{JaegerHaas2004,Yildiz2012}, to two-dimensional (2d) turbulent moist Rayleigh-Bénard convection (RBC) flow. We use this RNN to predict the low-order statistics, such as the buoyancy and liquid water fluxes or fluctuation profiles of these quantities across the layer. Our present work extends a recent application of ESNs to two-dimensional dry Rayleigh-B\'{e}nard convection \cite{Pandey2020} in several points. (1) The physical complexity of turbulent convection is enhanced in the present moist case. This is due to the total water content, an additional active scalar field which comprises of vapor and liquid water contents. The total water content couples as an additional field to the original dry Boussinesq model for temperature and velocity. (2) The size of the convection domain is increased by a factor of 4 such that the necessary degree of data reduction is significantly higher. (3) Moist convection requires also the satisfying reproduction of new physical quantities which are derived from different turbulence fields, e.g. the cloud cover in or the liquid water flux across the layer. This can be considered as a firmer test of the capabilities of the ESN to model complex dynamics. Such quantities are of particular interest for larger-scale models of atmospheric circulation that include mesoscale convection processes in form of parameters or minimal models \cite{grabowski_crcp_1999,grabowski_coupling_2001}. (4) Finally, the hyperparameters of the ESN, in particular the spectral radius $\rho(W^r)$ of the reservoir matrix $W^r$, has been tested in more detail (exact definitions follow). The grid search in our work thus sums up to a total of more than 1800 different hyperparameter sets.  

Echo state networks have recently received renewed attention for their capability of equation-free modeling of several chaotic systems, such as of the Lorenz96 model~\cite{vlachas2020} or the one-dimensional partial differential Kuramoto-Sivashinsky equation~\cite{lu_reservoir_2017,pathak2018}. Furthermore, low-order flow statistics in 2d dry Rayleigh-Bénard convection with ${\rm Ra}=10^7$ have been successfully reproduced using an ESN that trains the most energetic time coefficients of a Karhunen-Loéve expansion (or proper orthogonal decomposition) of the convection fields \cite{PhilHolmesTextbook}. This latter step can be considered as an autoencoder that provides training and test data for the ESN which cannot take the data directly, even for the present 2d case.  

A second popular implementation of RNNs, which we want to mention here for completeness, are {\em long short-term memory networks} (LSTM) \cite{SchmidhuberHochreiter1997} which have been also applied to fluid mechanical problems, such as the dynamics of Galerkin models of shear flows in \cite{srinivasan2019}. These models also demonstrated to capture the longer-term time dependency and low-order statistics of important turbulent transport properties well (see also \cite{pandey_perspective2020} for a direct comparison). An acceleration of the training, which requires the backpropagation of the errors through the whole network in contrast to ESNs, were obtained recently with Momentum LSTMs that apply the momentum extension of gradient descent search of the cost function minimum to determine the weights at the network nodes \cite{Tan2020}.     

In this work, a moist Rayleigh-Bénard convection model with moist Rayleigh number ${\rm Ra}_{\rm M} \simeq 10^8$ and Prandtl number $\text{Pr}=0.7$ is considered as an example case. We choose a 2d domain $\Omega=L\times H$ with aspect ratio $A=L/H=24$. Here $L$ is the horizontal length and $H$ the height of the simulation domain.  The number of grid points was chosen as $N_x\times N_y = 7200\times 384$. The data are obtained by direct numerical simulations (DNS) which apply a spectral element method \cite{Fischer1997,Scheel2013,nek5000}. Comprehensive studies of further data sets with different parameters, such as Rayleigh numbers, to study the generalization properties will be presented elsewhere.  Our intention is to demonstrate the capability of the ESN to deliver reliable low-order statistics for a turbulent convection flow with phase changes.

The manuscript is structured as follows. The next section introduces the moist RBC model and the central ideas of ESNs. Then the generation and analysis of the DNS data, including a brief description of the numerical scheme and the proper orthogonal decomposition (POD), is specified. Finally the results of our machine learning approach to moist turbulent convection will be discussed in detail. In section V we summarize our results and provide a conclusion and an outlook. 

%------------------------------------------------------------------------------------------------------------
%                 METHODS
%------------------------------------------------------------------------------------------------------------
\section{Methods}
\label{sec:materials_methods}
\subsection{Moist Rayleigh-Bénard Convection Model}
\label{subsec:mrbc}
\noindent
We now briefly review our model for moist Rayleigh-Bénard convection in two spatial dimensions. A detailed derivation can be found in~\cite{pauluis_idealized_2010, Weidauer_2010, schumacher_pauluis_2010}. The framework is based on the mathematically equivalent formulation by  Bretherton~\cite{bretherton_theory_1987,bretherton_theory_1988}. Similar simplified models of moist convection with precipitation have been developed by Smith and Stechmann \cite{Smith2017} and Vallis {\it et al.} \cite{VallisJFM2019}. Evaporative cooling and buoyancy reversal effects were discussed for example by Abma {\it et al.} \cite{AbmaJAS2013}.   

The buoyancy $B$ in atmospheric convection is given by~\cite{emmanuel1994}
\begin{align}
	\label{eq:buoyancy_general}
	B = -g\frac{\rho(S,q_v,q_l,q_i,p)-\overline{\rho}}{\overline{\rho}}
\end{align}
with the gravity acceleration $g$, a mean density $\overline{\rho}$, the pressure $p$, the entropy $S$ and the contents of water vapor $q_v$, liquid water $q_l$ and ice $q_i$. We consider warm clouds only, i.e. $q_i = 0$ and assume local thermodynamic equilibrium. From the latter assumption, it follows that no precipitation is possible and the number of independent variables in eq. (\ref{eq:buoyancy_general}) reduces to three. By introducing the total water content $q_T = q_v + q_l$ the buoyancy can be expressed as $B(S,q_T,p)$. In the Boussinesq approximation, pressure variations about a hydrostatic equilibrium profile are small, such that the buoyancy becomes $B(S,q_T,y)$ with the vertical spatial coordinate $y$ for the present 2d case. Furthermore, the convection layer is near the vapor-liquid phase boundary. The buoyancy can then be expressed as a piecewise linear function of $S$ and $q_T$ on both sides of the saturation line. This step preserves the discontinuity of the first partial derivatives of $B$ and therefore the physics of a first-order phase transition. The advantage of this formulation is that locally the saturation state of the air can be determined. In the last step, we substitute the linear combinations of $S$ and $q_T$ on both sides of the phase boundary by a dry buoyancy $D$ and a moist buoyancy $M$. Consequently the buoyancy field $B$ is given by \cite{pauluis_idealized_2010}
\begin{align}
	\label{eq:buoyancy_mrbc}
	B(x,y,t)=\max\left(M(x,y,t),D(x,y,t)-N_s^2y\right)
\end{align}  
where the fixed Brunt-V\"ais\"al\"a frequency $N_s = \sqrt{g(\Gamma_u - \Gamma_s)/T_{\rm ref}}$ is determined by the lapse rates of the saturated and unsaturated moist air, $\Gamma_s$ and $\Gamma_u$, and a reference temperature $T_{\rm ref}$. An air parcel at height $y$ and time $t$ is unsaturated if $M(x,y,t)<D(x,y,t)-N_s^2y$ and saturated if $M(x,y,t)> D(x,y,t)-N_s^2y$.
Note that the newly introduced dry buoyancy field $D$ is proportional to the liquid water static energy and the moist buoyancy field $M$ to the moist static energy. As in dry Rayleigh-Bénard convection with Dirichlet boundary conditions for the temperature, the static diffusive profiles $\overline{D}(y),\overline{M}(y)$ are vertically linear
%------------------------------------------------------
\begin{align}
\overline{D}(y)&=D_0 + \frac{D_H-D_0}{H}y\\
\overline{M}(y)&=M_0 + \frac{M_H-M_0}{H}y
\end{align}
%------------------------------------------------------
where $D_0$, $M_0$  and $D_H$, $M_H$ are the imposed values of $D$, $M$ at the bottom ($y=0$) and top ($y=H$) of the computational domain. Here, $D_0=M_0$.
The governing equations of the moist Boussinesq system are given by
%------------------------------------------------------
\begin{align}
	\label{eq:Navier-Stokes}
	\frac{d\mathbf{v}}{dt}&=-\nabla \tilde p +\nu\nabla^2\mathbf{v}+B\left(D,M,y\right)\hat{\mathbf{e}}_y\\
	\label{eq:MassBalance}
	\nabla\cdot \mathbf{v}&=0\\
	\frac{dD}{dt}&=\kappa\nabla^2 D\\
	\label{eq:Moist-Advection-Diffusion}
	\frac{dM}{dt}&=\kappa\nabla^2 M.
\end{align}
%------------------------------------------------------
Here $\mathbf{v}=(v_x(x,y,t),v_y(x,y,t))^T$ is the two-dimensional velocity field, $\tilde p=p/\rho_{\rm ref}$ the kinematic pressure, $\nu$ the kinematic viscosity and $\kappa$ the scalar diffusivity. The term $d/dt=\partial/\partial t+(\mathbf{v}\cdot \nabla)$ is the material derivative. This idealized model describes the formation of warm, non-precipitating low clouds in a shallow layer up to a depth of $\sim 1$km. The assumptions, which are made here, hold for example to a good approximation over the subtropical oceans.

The problem is made dimensionless using length scale $[x,y]=H$, buoyancy scale $[B] = M_0-M_H$, and (free-fall) time scale $[t] = \sqrt{H/(M_0-M_H)}$. The characteristic velocity scale follows by $[v_x,v_y]=\sqrt{(M_0-M_H)H}$. Four dimensionless numbers can be identified: the Prandtl number, dry Rayleigh number and moist Rayleigh number are given by 
%------------------------------------------------------
\begin{align}
	\label{eq:Prandtl}
	\text{Pr} &= \frac{\nu}{\kappa}\\
	{\rm Ra}_{\rm D} &=\frac{\left(D_0-D_H\right) H^3}{\nu \kappa}\\
	{\rm Ra}_{\rm M}  &=\frac{\left(M_0-M_H\right) H^3}{\nu \kappa}\,.
\end{align}
%------------------------------------------------------
An additional parameter arises from the additional phase changes, the dimensionless form of (\ref{eq:buoyancy_mrbc})
%------------------------------------------------------
\begin{equation}
	{\rm CSA} =\frac{N_s^2H}{M_0-M_H}\,.
	\label{eq:CSA}
\end{equation}
%------------------------------------------------------
The condensation in saturated ascent (CSA) controls the amount of latent heat an ascending saturated parcel can release on its way to the top. The saturation condition \eqref{eq:buoyancy_mrbc} implies that liquid water is immediately formed at a point in space and time when $M>D-N_s^2 y$. There is no supersaturation considered in this model and the liquid water content field $q_l$ and thus the clouds are given by 
%------------------------------------------------------
\begin{equation}
	q_l(x,y,t) =M(x,y,t)-(D(x,y,t)-N_s^2 y)\,.
	\label{eq:ql}
\end{equation}
%------------------------------------------------------
Note that in this formulation, $q_l$ can become negative, as it is a measure of the degree of saturation. In a nutshell, $q_l<0$ stand thus for a liquid water deficit. When the atmosphere is saturated, $q_l\ge0$ and the conventional liquid water content is retained.
 
Here we study the case of $D_0>D_H$ and $M_0>M_H$. Both fields are linearly unstable. For the case of a {\em conditionally unstable} moist layer with $D_0\le D_H$ we refer to refs. \cite{bretherton_theory_1987,bretherton_theory_1988} or \cite{PauluisSchumacherPNAS2011}.

%------------------------------------------------------
\subsection{Reservoir Computing}
\label{subsec:rc}

Reservoir computing (RC) is a special type of RNN implementation. Contrary to standard feed forward networks, neurons in the hidden layers of RNN are recurrently connected to each other. In this way, RNNs have a similar architecture to biological brains and are said to posses an internal memory as cyclic connections allow information to stay inside the network for a certain amount of time before fading out, known as the {\em echo state property}. Yet the training of such RNNs is exceedingly difficult, as common training schemes like the back propagation through time struggle with fading error gradients, slow convergence~\cite{Jaeger2005ATO} and bifurcations~\cite{doya_bifurcations_2000}. An alternative training method was proposed by Jaeger~\cite{jaeger__2001} and in an independent work by Maass~\cite{maass_real-time_2002}. Their idea, which is now known as reservoir computing~\cite{LukoseviciusJaeger2009}, was to train the weights of the output layer only, which connect the RNN, denoted to as the {\em reservoir}, to the output neurons. The weights of the input layer as well as the internal reservoir weights should be initialized at random and then kept constant. This training procedure reduces the computational costs for training significantly and shifts the focus to an adequate initialization of the input and reservoir  weight matrix $W^r$ (which is an adjacency matrix in network theory). While Jaeger's approach is known as ESN, Maass' framework is called {\em liquid state machine}. They differ in their field of origin, as the ESN stems from the field of machine learning and the liquid state machine from computational neuroscience. We will stick to the ESN formulation, but note that RC refers to the concept mentioned above and summarizes both models.

Despite their simple training scheme, ESNs have been said to tackle many tasks, from forecasting closing prices of stock markets~\cite{Lin2008} to estimating the life span of a fuel cell~\cite{morando2013}. Especially its application to dynamical systems shows great promise. For instance it has been demonstrated that the dynamics of two of the three degrees of freedom of the R\"ossler system can be inferred from the evolution of the third one~\cite{lu_reservoir_2017}. Further, the Lyapunov exponents of the dynamical system that a trained ESN represents have been shown to match the exponents of the data generating system~\cite{pathak_using_2017}. 

%------------------------------------------------------
\begin{figure}
\includegraphics[width = 200pt]{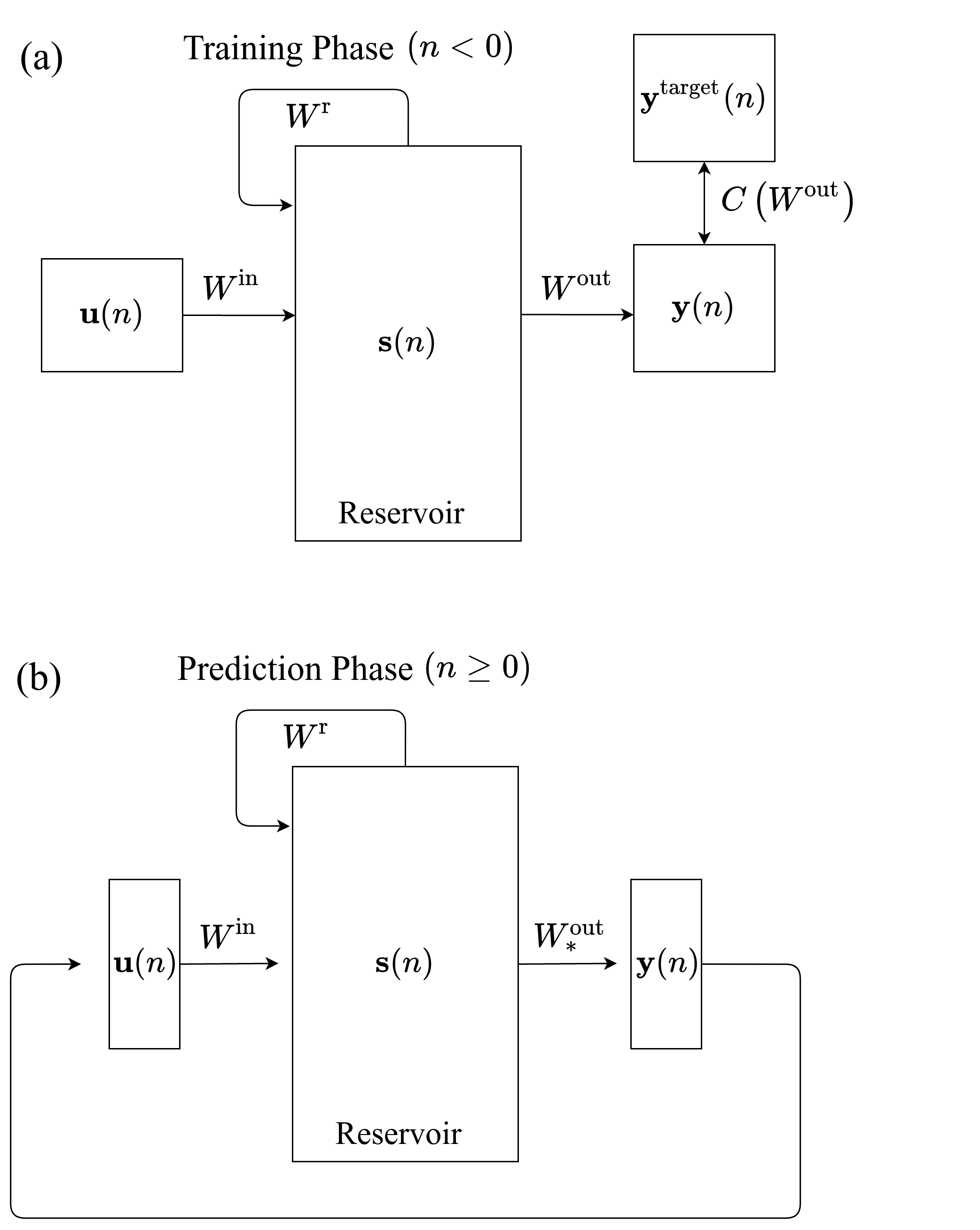}
\caption{Sketch of the echo state network in the training phase (a) for time steps $n<0$ and the prediction phase (b) for time steps $n\ge0$.}
\label{fig:rcm_concept}
\end{figure}
%------------------------------------------------------
Figure \ref{fig:rcm_concept} shows the concept and components of the ESN, for the training phase in panel (a) and for the prediction phase in panel (b). The input $\mathbf{u}(n)\in \mathbb{R}^{\rm N_{\rm in} }$ at a time instance $n$ as well as a constant scalar bias $b=1$ are passed to the reservoir via the input weight matrix $W^{\rm in}\in \mathbb{R}^{\rm N_{\rm r}\times (1+\rm N_{\rm in})}$. The weighted input contributes to the dynamics of the reservoir state $\mathbf{s}\in \left[-1,1\right]^{\rm N_{\rm r}}$ at time $n$ which is given by
%------------------------------------------------------
\begin{align}
	\label{eq:rcm_activation}
	\mathbf{s}(n)&= (1-\gamma)\mathbf{s}(n-1) \nonumber\\
	&+ \gamma\tanh\left[W^{ \rm in}\left [b;\mathbf{u}(n)\right] + W^{\rm r} \mathbf{s}(n-1)\right].
\end{align}
%------------------------------------------------------
Here $\left[b;\mathbf{u}(n)\right]$ stands for the vertical concatenation of the scalar bias and the input \footnote{In some cases the bias $b=0$ and $\mathbf{u}(n-1)$ are used in eq. \eqref{eq:rcm_activation}.}. This update rule comprises external forcing by the inputs $\mathbf{u}(n)$ as well as a self-interaction with the past instance $\mathbf{s}(n-1)$. The reservoir weight matrix $W^{\rm r}\in \mathbb{R}^{\rm N_{\rm r}\times \rm N_{\rm r}}$ blends the state dimensions, while the nonlinear hyperbolic tangent $\tanh\left(\cdot\right)$, applied to each component of its argument vector, is the nonlinear activation function of the neurons in this model. The leaking rate $\gamma\in \left(0,1\right]$ moderates the linear and nonlinear contributions and assures that the state is confined to $\left[-1,1\right]^{N_{\rm r}}$. As mentioned above, the existence of echo states, i.e., states that are purely defined by the input history, is crucial. An ESN is said to include such echo states when two different states $\mathbf{s}(n-1)$, $\mathbf{s}'(n-1)$ converge to the same state $\mathbf{s}(n)$, provided the same input $\mathbf{u}(n)$ is given and the system has been running for many iterations $n$~\cite{jaeger__2001}. Therefore, the first few state iterations are considered as a reservoir washout, even for a reservoir with echo state property. After this transient phase, the updated state is concatenated with the bias and the current input to form the extended reservoir state $\tilde{\mathbf{s}}(n)=\left[b;\mathbf{u}(n);\mathbf{s}(n)\right]$. Finally, $\tilde{\mathbf{s}}$ is mapped via the output matrix $W^{\rm out}\in \mathbb{R}^{\rm N_{\rm in}\times \left(1+\rm N_{\rm in}+\rm N_{\rm r}\right)}$ to form the reservoir output $\mathbf{y}(n)\in \mathbb{R}^{\rm N_{\rm in}}$
%------------------------------------------------------
\begin{equation}
	\label{eq:rcm_out}
	\mathbf{y}(n)=\mathrm{W}^{ \rm out}\tilde{\mathbf{s}}(n).
\end{equation}
%------------------------------------------------------
For our application the output dimension matches the input dimension $\rm N_{\rm in}$, which generally does not need to be the case.

Before the ESN can be used in the prediction phase, as sketched in Fig. \ref{fig:rcm_concept}(b), the elements of $W^{\rm out}$ have to be computed first. This process is known as training phase of this supervised machine learning algorithm. Only when the reservoir is properly trained it will produce reasonable output. A set of $n_{\rm train}$ training data instances $\{\mathbf{u}(n),\mathbf{y}^{\rm target}(n)\}$ (where $n= -n_{\rm train}, -(n_{\rm train}-1), ..., -1$) needs to be prepared. The target output $\mathbf{y}^{\rm target}(n)$ represents the desired output the ESN should produce for the given input $\mathbf{u}(n)$. The reservoir state $\mathbf{s}$ is computed for all inputs in the training data set and assembled into a mean square cost function with an additional $L_2$ regularization $C\left(W^{\rm out}\right)$ which is given by  
%------------------------------------------------------
\begin{align}
	C\left(W^{\rm out}\right) &= \frac{1}{n_{\rm train}}\sum_{n = -n_{\rm train}}^{-1} \left(W^{\rm out}\tilde{\mathbf{s}}(n)-\mathbf{y}^{\rm target}(n)\right)^2 \nonumber\\
	&+ \beta \sum\limits_{i=1}^{N_{\rm in}}\|w_i^{\rm out}\|_2^2,
	\label{eq:rcm_train1}
\end{align}
%------------------------------------------------------
and has to be minimized corresponding to 
%------------------------------------------------------
\begin{align}	
	W_{\ast}^{\rm out}& = \arg\min C\left(W^{\rm out}\right)\,.
	\label{eq:rcm_train2}
\end{align}
%------------------------------------------------------
Here $w_i^{\rm out}$ denotes the $i$-th row of $W^{\rm out}$ and $\|\cdot \|_2$ the $L_2$ norm. Equations \eqref{eq:rcm_train1} and \eqref{eq:rcm_train2} are known as ridge regression with the ridge regression parameter $\beta$. The last term in \eqref{eq:rcm_train1} suppresses large values of the rows of $W^{\rm out}$, which could inadvertently amplify small differences of the state dimensions in (\ref{eq:rcm_out}). This well known regression problem is solved by the fitted output matrix
%------------------------------------------------------
\begin{eqnarray}
\label{eq:rcm_fitted_wout}
W_{\ast}^{\rm out} = Y^{\rm target}S^{\rm T}\left(SS^{\rm T}+\beta {\rm Id}\right)^{-1}
\end{eqnarray}
%------------------------------------------------------
where $\left(\cdot\right)^{\rm T}$ denotes the transposed and ${\rm Id}$ the identity matrix. $Y^{\rm target}$ and $S$ are matrices where the $n$-th column is the target output $\mathbf{y}^{\rm target}(n)$ and the extended reservoir state $\tilde{\mathbf{s}}(n)$, respectively. 

As the output weights are the only parameters that are trained, RC is computationally inexpensive. However, the algebraic properties of the initially randomly generated matrices $W^{\rm in}$ and $W^{\rm r}$ are hyperparameters which have to be tuned beforehand. In our approach we draw the elements of $W^{\rm in}$ from a uniform distribution in $\left[-0.5,0.5\right]$ and impose no further restrictions on this matrix. For the generation of the reservoir weights in $W^{\rm r}$ it has been reported that the proportion of non-zero elements, i.e., the reservoir density D and the spectral radius $\varrho\left(W^{\rm r}\right)$, i.e., the largest absolute eigenvalue of $W^{\rm r}$ are crucial parameters that determine whether the desired echo state property holds~\cite{lukosevicius_practical_2012}. We choose a sparse reservoir ($\text{D}< 1$) with few internal node connections and draw the elements from a uniform distribution in $\left[-1,1\right]$. We then normalize $W^{\rm r}$ by its largest absolute eigenvalue and multiply it with the desired spectral radius $\varrho\left(W^{\rm r}\right)$. This scaling approach, initially proposed by Jaeger~\cite{Jaeger2005ATO}, has established itself as one of the standard ways~\cite{morando2013,lu_reservoir_2017, pathak2018,Pandey2020} to control the spectral radius. Nevertheless, other procedures have been proposed~\cite{Yildiz2012,strauss_design_2012}. In addition, the size of the reservoir $\rm N_{\rm r}$ is a hyperparameter. Usually $\mathbf{s}$ should be a high-dimensional extension of the inputs $\mathbf{u}$, so that $\rm N_{\rm r}\gg \rm N_{\rm in}$ is satisfied. Moreover, we consider the leaking rate $\gamma$ and ridge regression parameter $\beta$ as further quantities that have to be adjusted to our data. 

%------------------------------------------------------------------------------------------------------------
%                 APPLICATION
%------------------------------------------------------------------------------------------------------------
\section{Echo state network for 2d Moist Convection}
\label{sec:Application}
\subsection{Direct Numerical Simulations}
\label{subsec:dns}
\begin{figure}[!htbp]
	\includegraphics[width = 240pt]{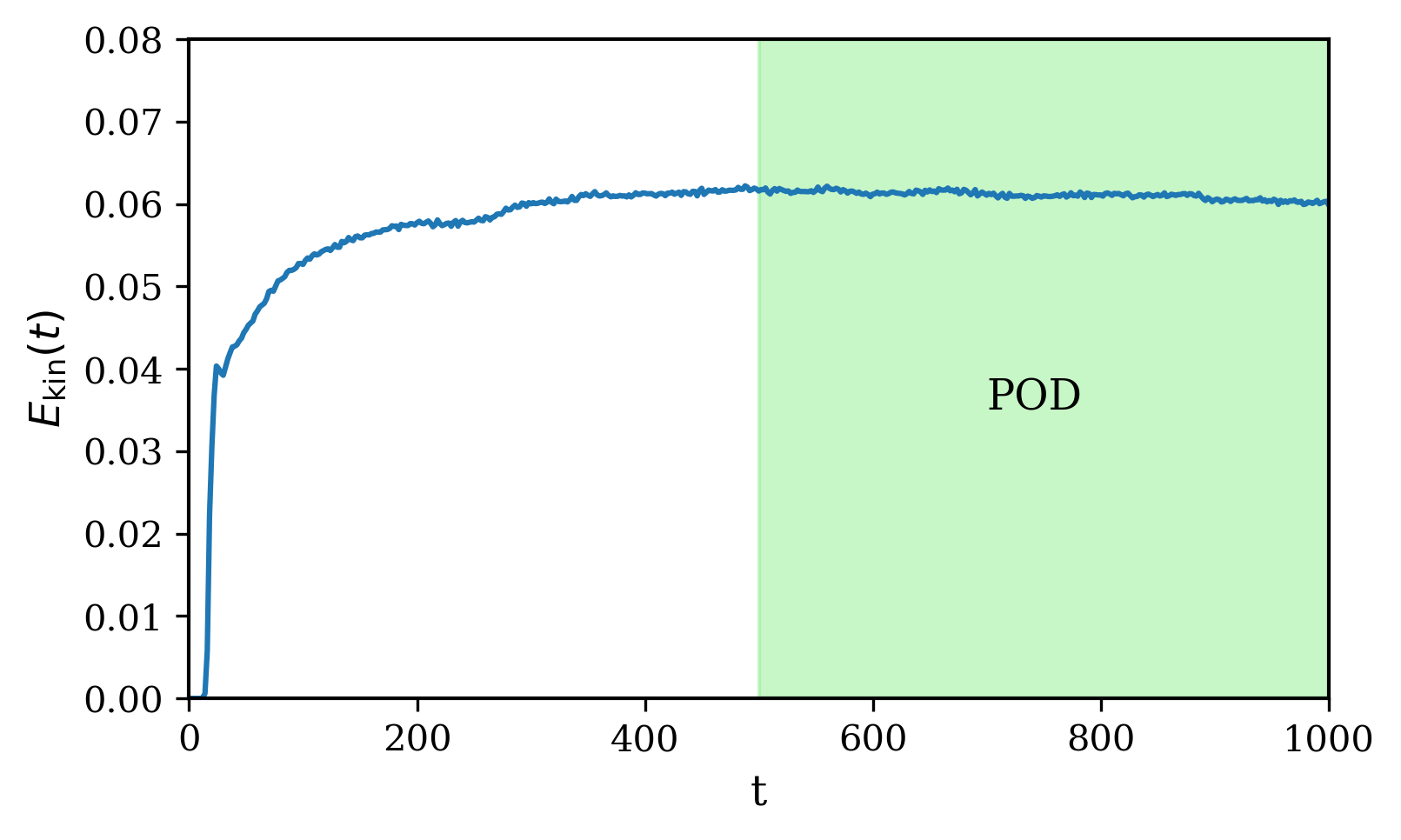}
	\caption{Turbulent kinetic energy $E_{\rm kin}(t) = \langle v_x^2 + v_y^2\rangle_{x,y}$ of the moist Rayleigh-Bénard flow versus time $t$. After an initial transient the values such as those of $E_{\rm kin}(t)$ become statistically stationary. In the statistically stationary regime,  $2000$ snapshots, each separated by $\Delta t=0.25$, are analyzed by a POD (see Sec. \ref{subsec:rcm}). }
	\label{fig:DNS_sss}
\end{figure}

\noindent
DNS using the spectral element solver Nek5000~\cite{Fischer1997,Scheel2013,nek5000} were conducted to solve the two-dimensional moist Rayleigh-Bénard system (\ref{eq:Navier-Stokes})-(\ref{eq:Moist-Advection-Diffusion}) in a domain $\Omega = L\times H$ with aspect ratio $ A=L/H=24$. The Rayleigh numbers are ${\rm Ra}_{\rm D} = 2\cdot 10^8$, ${\rm Ra}_{\rm M} = 4\cdot 10^8$. The Prandtl number is ${\rm Pr} = 0.7$ representing moist air. The additional parameter is set to ${\rm CSA = 0.3}$. In the vertical direction $y$, Dirichlet boundary conditions are imposed for both buoyancy fields at top and bottom in combination with free-slip boundary conditions for the velocity field. Periodic boundaries are set for all fields in the horizontal direction. We chose a spatial resolution of $N_x\times N_y = 7200\times 384$ grid points and a time step size of $5.0\cdot 10^{-4}$.  This setup corresponds to an absolutely unstable atmosphere, i.e. where both unsaturated and saturated air are unstable w.r.t. vertical displacements. The initial conditions are small perturbations around the diffusive equilibrium state $\overline{M}(y)$ and $\overline{D}(y)$, which result in turbulent convection. The flow statistics relaxes into a statistically stationary state (see Fig. \ref{fig:DNS_sss}) which provides training and test data for further processing.
%------------------------------------------------------
%------------------------------------------------------

\subsection{Data reduction by POD}
\label{subsec:rcm}
\noindent
We sample a total of $n_s = 2000$ snapshots of $v_x$, $v_y$, $M$ and $D$ at a time interval $\Delta t=0.25$ in the statistically stationary regime. The original DNS data snapshots have been sampled at a constant time interval  such that we stick to  constant time steps throughout this work, including the subsequent RC model. Furthermore, the data are spectrally interpolated on a uniform grid with a resolution of $N_x'\times N_y' = 640\times 80$ points from the originally unstructured element mesh. They are decomposed into temporal mean and fluctuations subsequently,
%------------------------------------------------------
\begin{eqnarray}
	v_x(x,y,t) &=& \langle v_x\rangle_{t}(x,y) + v_x^{\prime}(x,y,t)\\
	\label{eq:reynolds_ux}
	v_y(x,y,t) &=& \langle v_y\rangle_{t}(x,y) + v_y^{\prime}(x,y,t)\\
	\label{eq:reynolds_uy}
	D(x,y,t) &=& \langle D\rangle_{t}(x,y) + D^{\prime}(x,y,t)\\
	\label{eq:reynolds_D}
	M(x,y,t) &=& \langle M\rangle_{t}(x,y) + M^{\prime}(x,y,t)\\
	\label{eq:reynolds_M}
	q_l(x,y,t) &=& \langle q_l\rangle_{t}(x,y) + q_l^{\prime}(x,y,t)\,.
	\label{eq:reynolds_ql}	
\end{eqnarray}
%------------------------------------------------------
%------------------------------------------------------
\begin{figure}[htpb]
	\includegraphics[width = 240pt]{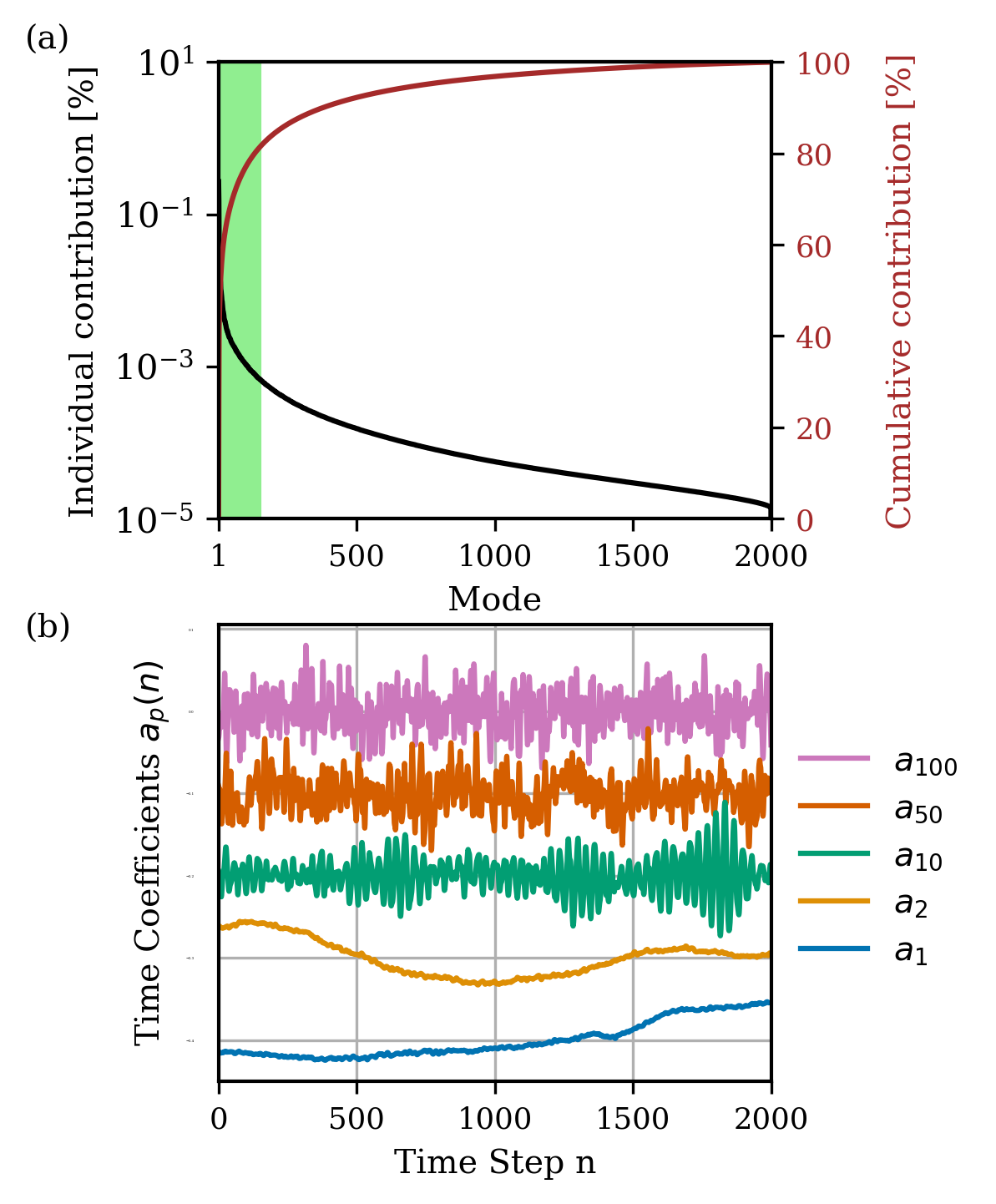}
	\caption{Eigenvalue spectrum of the POD mode obtained from the analysis of 2000 snapshots. (a) Individual and cumulative contribution of each mode.  The shaded region indicates the first $\rm N_{\rm POD}=150$ modes which capture $81\%$ of the total energy of the original snapshot data. (b) Time coefficients $a_p(n)$ for the 1st, 2nd, 10th, 50th, and 100th mode are shown. The first coefficients show a slow variation compared to higher coefficients. Time series are shifted with respect to each other for better visibility.}
	\label{fig:pod_spectrum}
\end{figure}
%------------------------------------------------------
A grid dimension of $640\times 80$ in terms of ESN input dimensions ${\rm N_{\rm in}}$ is still too big. We therefore follow the approach in~\cite{Pandey2020} and introduce an intermediate step of data reduction before handing the data to the ESN. We make use of the periodicity and expand the data in a Fourier series in the horizontal $x$-direction and take the Karhunen-Loéve expansion, also known as POD of our data. In particular we choose the snapshot method \cite{Sirovich1987} which decomposes the $k$-th component of a vector field $\mathbf{g}(x,y,t)$ into
%------------------------------------------------------
\begin{align}
	g_k(x,y,t)&= \sum\limits_{p = 1}^{\rm n_{\rm s}} \sum_{n_x =-N_x'/2}^{N_x'/2} a_{p,n_x}(t) \Phi_{k, n_x}^{(p)}(y) \exp{\left(i\frac{2\pi n_x x}{L}\right)} \nonumber \\
	&= \sum\limits_{p = 1}^{ \rm n_{\rm s}} a_{p}(t) \Phi_{k}^{(p)}(x,y).
	\label{eq:pod}
\end{align}
%------------------------------------------------------
Here $a_{p}(t)$ and $\Phi_{k}^{(p)}(x,y)$ are the $p$-th time coefficient and the corresponding spatial POD mode respectively.  In our approach we take the POD of $\mathbf{g}=(v_x',v_y',D',M')^T$. The POD spectrum of the turbulent convection data can be seen in Fig. \ref{fig:pod_spectrum}(a). The eigenvalues of the covariance matrix fall off quickly and we therefore truncate the POD expansion at $\rm N_{\rm POD}= 150 \ll \rm n_{\rm s}$ and include only the most energetic modes (green shaded region). These capture $81 \%$ of the total energy of the original data. In Fig. \ref{fig:pod_spectrum}(b) the time coefficients $a_p(n)$ are shown for $p=1,2,10,50$, and 100 for all POD time steps. The first time coefficients ($a_1$ to $a_{10}$) posses only few temporal features opposed to higher coefficients. This range of active scales is inherent to turbulence as kinetic energy of large-scale motion is transferred to small eddies down to the Kolmogorov scale.  Moreover, the influence of the additional nonlinearity due to the phase changes impacts the dynamics, as the first coefficients varied more in the dry RBC case with aspect ratio $6$ at ${\rm Ra = 10^{7}}$~\cite{Pandey2020}. This is one order of magnitude below our Rayleigh number values. Nevertheless our RC model will receive values for all $\rm N_{\rm POD}$ coefficients and therefore for a wide range of temporal frequencies.  We note here that the design of the  RC model could be adapted to the different frequencies in future approaches. 

\begin{figure*}[htpb]
		\includegraphics[width = 482pt]{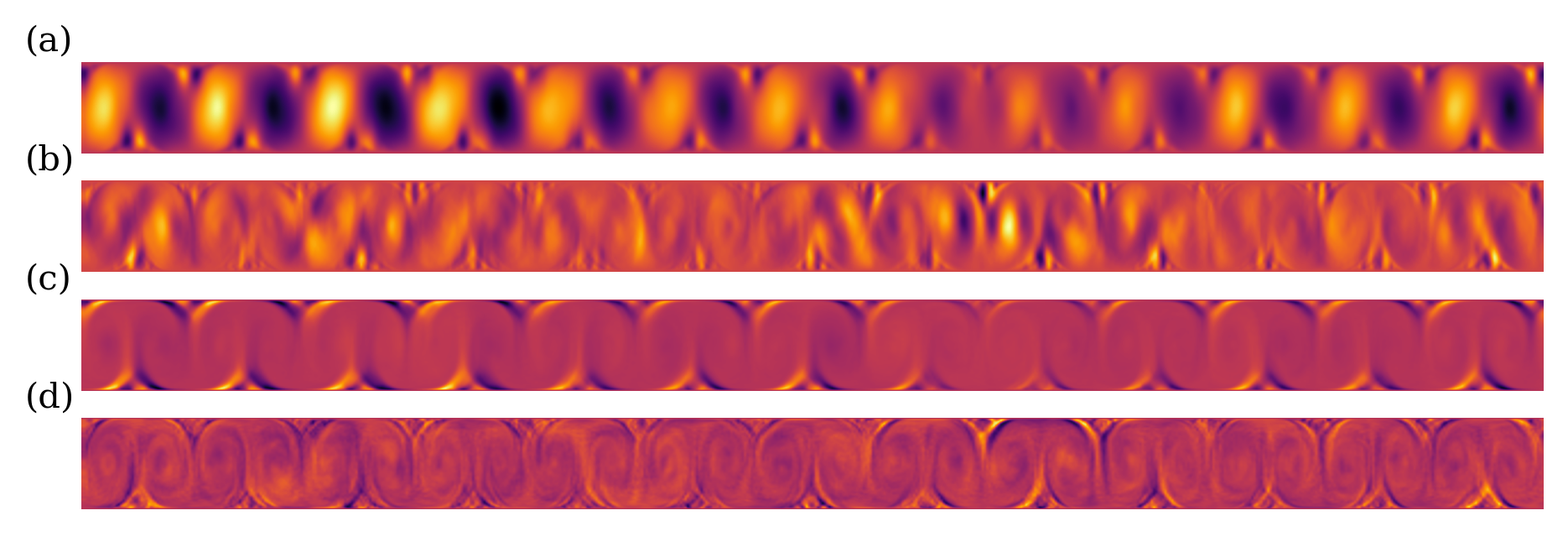}
	\caption{Spatial structure of two POD modes for $v_y$: (a) $\Phi_{2}^{(1)}(x,y)$, (b) $\Phi_{2}^{(50)}(x,y)$ and the moist buoyancy field $M$: (c) $\Phi_{4}^{(1)}(x,y)$, (d) $\Phi_{4}^{(50)}(x,y)$. For visualization purposes the aspect proportions do not match the actual aspect ratio of $A=24$.}
	\label{fig:pod_modes}
\end{figure*}

Figure \ref{fig:pod_modes} shows the spatial modes $\Phi_{2}^{(1)}$, $\Phi_{2}^{(50)}$ and $\Phi_{4}^{(1)}$, $\Phi_{4}^{(50)}$. 
%------------------------------------------------------
%------------------------------------------------------
We limit ourselves to the last $1400= {\rm n}_{\rm train} + {\rm n}_{\rm test}$ time instances of our data and use the first 150 time coefficients $\mathbf{a}(n)= \left(a_1(n),a_2(n),...,a_{150}(n)\right)^T$ as the input for the ESN. The first ${\rm n}_{\rm train}=700$ snapshots are assembled into a training data set
\begin{align}
\{ \mathbf{u}(n), \mathbf{y}^{\rm target}(n) \}=\{ \mathbf{a}(n), \mathbf{a}(n+1) \}
\end{align}
with $-{\rm n}_{\rm train}\le n\le -1$. The training data span 175 free-fall time units $T_f$ that correspond to 61 eddy turnover times. This time scale is given by $\tau_{\rm eddy}=H/u_{\rm rms}\approx 2.9 T_f$ with $u_{\rm rms}=\langle u_x^2+u_y^2\rangle^{1/2}_{x,y,t}$. The ESN is trained to predict the time instance $\mathbf{a}(n+1)$ when given the POD time coefficients $\mathbf{a}(n)$ at the last time step as input. We use the first $46$ time steps to initialize the reservoir. Using eq. (\ref{eq:rcm_fitted_wout}) we compute $W^{\rm out}$, which can then be used for prediction. In the prediction phase, we give the initial input $\mathbf{u}(0) = \mathbf{a}(0)$ to the reservoir and redirect the output to the input layer (see Fig. \ref{fig:rcm_concept}(b)) such that
\begin{align}
	\mathbf{u}(n) &= \mathbf{y}(n-1) \qquad n = 1,..., {\rm n}_{\rm test}-1
\label{eq:rcm_couple}	
\end{align}
with ${\rm n}_{\rm test}=700$. This coupling creates an autonomous system that generates new outputs without providing external inputs. Contrary to the teacher forcing approach, where at each time step the input is given by the actual time coefficients, this method is more suited for real world application. Finally, the outputs at each time step are gathered and validated by the last ${\rm n}_{\rm test}$ snapshots $\{\mathbf{y}^{\rm val}(n)\}=\{\mathbf{a}(n+1)\}$. 

\subsection{Training of ESN and choice of hyperparameters}
\noindent
We quantify the quality of ESN predictions, with the set of hyperparameters $\rm h = \{ \gamma,\beta,\rm N_{\rm r},{\rm D},\varrho(W^{\rm r})  \}$, by two types of measures. The mean square error ${\rm MSE}_{\rm h}$ of ESN output to the validation data
%------------------------------------------------------
\begin{align}
	{\rm MSE}_{\rm h}
	&= \frac{1}{{\rm n}_{\rm test}}\sum\limits_{n = 0}^{{\rm n}_{\rm test}}{\rm mse}(n)
\intertext{where}
	{\rm mse}(n)&=\frac{1}{\rm N_{\rm POD}}\sum\limits_{i = 1}^{{\rm N}_{\rm POD} } \left(y_i(n)-y^{\rm val}_i(n) \right)^2
	\label{eq:mode_mse}
\end{align}
is the mean square error at time step $n$ averaged over all ${\rm N}_{\rm POD}$ modes. Additionally, we take the physically more relevant normalized average relative error (NARE) as defined in~\cite{srinivasan2019} into account. For the moist buoyancy field $M$, it is for example given by
\begin{align}
	E_{\rm h}\left[\langle M\rangle_{x,t}\right] &= \frac{1}{C_{\max}}\int\limits_0^1 \Big|\langle M\rangle_{x,t}^{\rm ESN}(y) -\langle M\rangle_{x,t}^{\rm POD}(y)\Big|dy
	\label{eq:nare}
	\intertext{with}
	C_{\max}&=\frac{1}{2 \max_{y\in [0,1]}(|\langle M\rangle_{x,t}^{\rm POD}|)}.
\end{align}
The superscript defines whether the field was reconstructed with $a_i(n)$ (POD) or $y_i(n)$ (ESN). It measures the integral deviation of the reconstructed line-time average profile $\langle \cdot \rangle_{x,t}$ of a specific field. We consider the three NAREs: $E_{\rm h}\left[\langle M\rangle_{x,t}\right]$, $E_{\rm h}\left[\langle q_l'^2\rangle_{x,t}\right]$ and $E_{\rm h}\left[\langle v_y'M'\rangle_{x,t}\right]$, that is the NARE of the total moist buoyancy field $M$, the liquid water content fluctuations $q_l'$, and the moist buoyancy flux fluctuations $v_y'M'$.

%------------------------------------------------------
\begin{table}[!hpbt]
	\begin{ruledtabular}
		\begin{tabular}{cc  cc }
			$\gamma$ &  $\beta$	          &$D$ 		& $\varrho(W^{\rm r})$\\
			\hline 
			$0.50$  &  $5\cdot 10^{-4}$  &$0.1$   &$0.00$   \\
			$0.60$   &  $5\cdot 10^{-3}$  &$0.2$   &$0.90$  \\
			$0.70$   &  $5\cdot 10^{-2}$  &$0.3$   &$0.91$  \\
			$0.80$   &  $5\cdot 10^{-1}$  &$0.4$   &$0.92$  \\
			$0.90$   &                                         &$0.5$   &$0.93$  \\
			$0.95$&                                            &$0.6$   &$0.94$  \\
			&                                           &$0.7$   &$0.95$  \\
			&          							    &               &$0.96$  \\                
			&          							    &               &$0.97$  \\
			&                 							&		        &$0.98$  \\
			&                 							&		        &$0.99$  \\
			&                 							&		        &$1.00$  \\
			
		\end{tabular}
	\end{ruledtabular}
	\caption{Range of the four hyperparameters upon which a grid search was conducted. For each of the $1848$ combinations, an ESN was trained and validated with the training and validation data set. The reservoir dimension $N_{\rm r}$ was set to $4000$ for all studies. The MSE and NARE measures were computed and evaluated to find the optimal parameter set $h^*$.}
	\label{tab:grid_search}
\end{table}
%------------------------------------------------------
\begin{table}[!hpbt]
	\begin{ruledtabular}
		\begin{tabular}{c cc cc }
			$\gamma^{*}$  &$\beta^{*}$  &$N_{\rm r}^{*}$ &$D^{*}$& $\varrho(W^{\rm r})^{*}$\\
			\hline
			$0.9$   &$5\cdot 10^{-1}$   &$4000$  &$0.1$  &$1.0$ \\
		\end{tabular}
		\vspace*{.5cm}
		\begin{tabular}{cccc}
			MSE($\rm h^*$)&  $E_{\rm h}\left[\langle M\rangle_{x,t}\right]$ & $E_{\rm h}\left[\langle q_l'^2\rangle_{x,t}\right]$ & $E_{\rm h}\left[\langle v_y'M'\rangle_{x,t}\right]$\\
			\hline
			$8.18\cdot 10^{-4}$ &$0.032$\% &$0.033$\% &$4.5$\% \\
		\end{tabular}
	\end{ruledtabular}
	\caption{Hyperparameter set $h^*$ that was chosen for the ESN setup and the associated errors, which this ESN run has produced. The results of the reservoir with these listed hyperparameters are presented in section \ref{sec:results}.  }
	\label{tab:hyperparam}
\end{table}
%------------------------------------------------------
\begin{figure*}[hptb!]
	\includegraphics[width = 14cm]{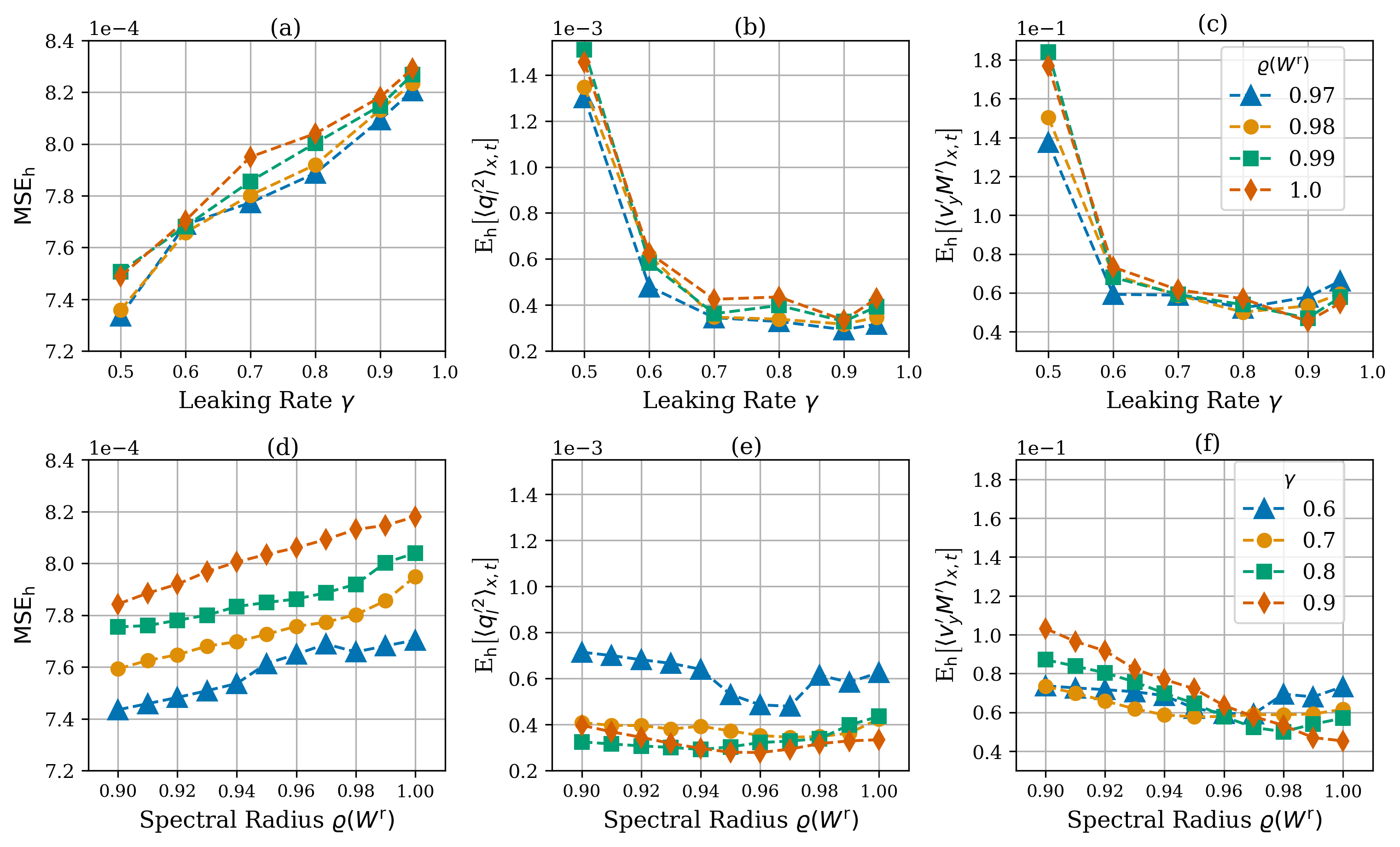}
	\caption{Representative profiles taken from the error landscape for the leaking rate $\gamma$ (a,b,c) and the spectral radius $\varrho(W^{\rm r})$ (d,e,f). The data are obtained by a grid search study (see Table \ref{tab:grid_search}). We find a systematic dependence for the two quantities in this parameter domain. Note the different magnitudes between single quantity-NARE in (b,e) and multiple quantity-NARE in panels (c,f). The legends in (c) and (f) hold also for panels (a,b) and (d,e), respectively.}
	\label{fig:grid_search}
\end{figure*}
%------------------------------------------------------

A grid search for the four quantities $\gamma$, $\beta$, $D$, $\varrho(W^{\rm r})$ was conducted in a suitable range (see Table \ref{tab:grid_search}), based on the results in~\cite{Pandey2020}. The reservoir size was fixed to $N_{\rm r}= 4000$ for all runs. The resulting ${\rm MSE}_h$ and NAREs were computed to find an adequate parameter setting. Figure \ref{fig:grid_search} shows ${\rm MSE}_h$, $E_{\rm h}\left[\langle q_l'^2\rangle_{x,t}\right]$, $E_{\rm h}\left[\langle v_y'M'\rangle_{x,t}\right]$ and their dependence on $\varrho\left(W^{\rm r}\right)$ and $\gamma$. We detect a systematic dependence of both, spectral radius and leaking rate, even when slightly changing a third parameter (see legends). Interestingly, as both parameters are increased, the mean square error increases as well, while the NARE either decrease or pass a local minimum. We emphasize that our grid search is only an excerpt of the much bigger error landscape. Further, we did not average over multiple random realizations which would be the basis for a more rigorous discussion of parameter dependencies. Nevertheless, a starting point of the discussion of the hyperparameter dependencies is as follows: as the spectral radius grows, the magnitude of the argument of the hyperbolic tangent builds up too. This will saturate the activation function, which in turn will act in an increasingly binary way since $\lim\limits_{x\rightarrow \pm\infty} \tanh(x) = \pm 1$. In this limit of a fully saturated activation function, eq. (\ref{eq:rcm_activation}) simplifies to
\begin{align}
	\mathbf{s}(n) \simeq (1-\gamma)\mathbf{s}(n-1) \pm \gamma \mathbf{1},
	\label{limit}
\end{align}
where $\mathbf{1}=(1,1,...,1)^T\in \mathbb{R}^{N_{\rm r}}$. This corresponds to a linear dependence of each reservoir state on its last instance plus the constant leaking rate $\gamma$ with stochastically changing sign, depending on the randomly generated weight matrices. As the leaking rate is increased towards unity, the memory effect is lost as well and the reservoir state is basically updated by the constant last term in \eqref{limit}. The resulting reservoir output will lead to increased mean square deviations from the varying ground truth signal. We thus speculate that such activation saturation is already satisfied for several reservoir state components at $\rho(W^r)\lesssim 1$ which in turn contribute to the increasing ${\rm MSE}_{\rm h}$ in panel (d) of Fig. \ref{fig:grid_search}. 

Finally, we chose the hyperparameter set  $h^{*}$, listed in Table \ref{tab:hyperparam} as it results in a minimum of $E_{\rm h}\left[\langle v_y'M'\rangle_{x,t}\right]$. The reason for settling with this measure is that it is susceptible to two ESN estimates. Quantities like $E_{\rm h}\left[\langle q_l'^2\rangle_{x,t}\right]$, which depend on only one ESN estimate, exhibit small values for many parameter settings (see Fig. \ref{fig:grid_search} (b),(e)).
%------------------------------------------------------------------------------------------------------------
%                 RESULTS
%------------------------------------------------------------------------------------------------------------
\section{Results for the moist RBC case}
\label{sec:results}
%------------------------------------------------------
\begin{figure*}[!hptb]
	\includegraphics[width = 14cm]{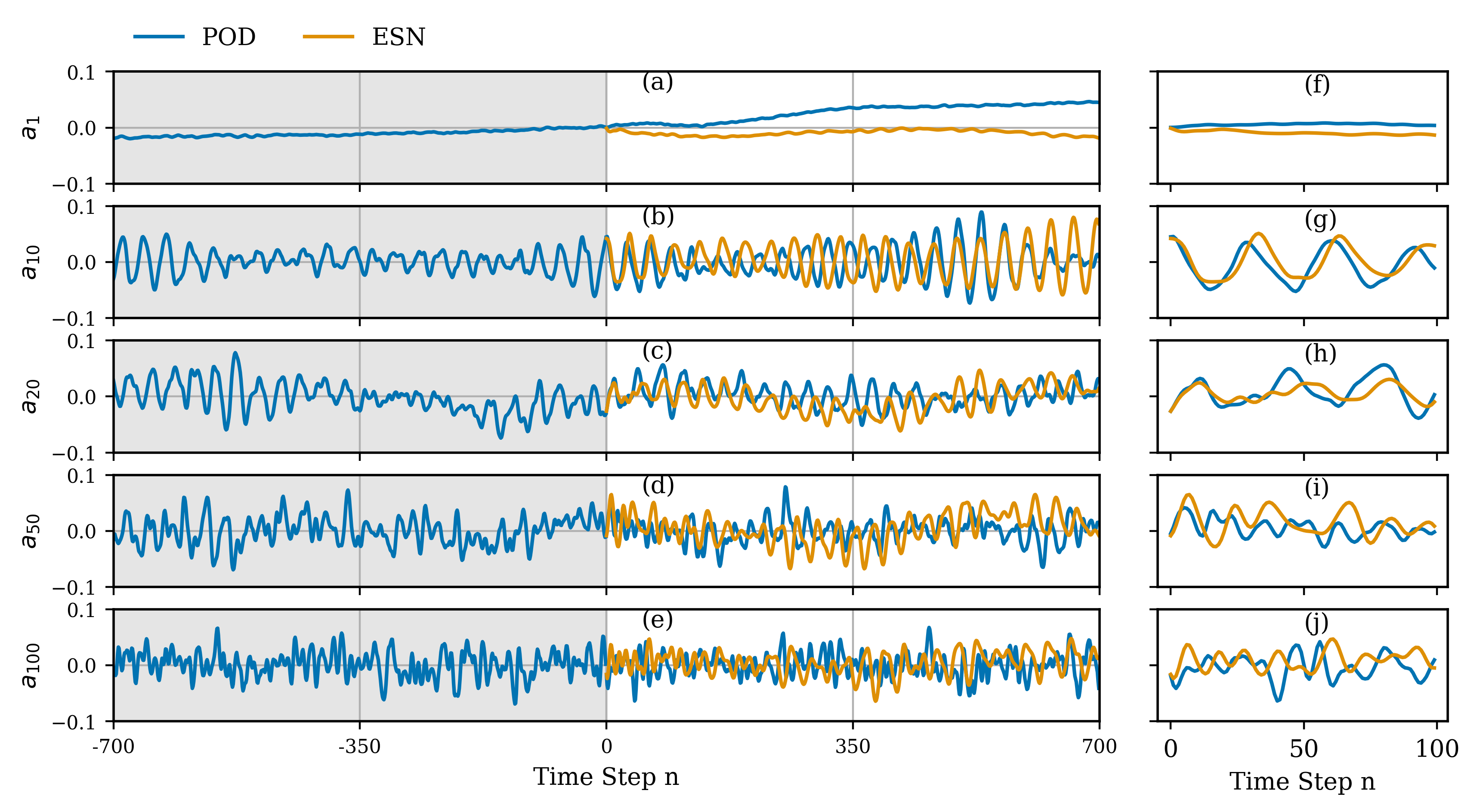}
	\caption{Time evolution of the POD time coefficients $a_i(t)$. The gray shaded area marks the training phase (reservoir output not shown). At the end of the training phase the prediction phase starts. The curves labeled POD stand for the ground truth of the evolution of the coefficient, while those labeled ESN are the network predictions. Panels (f)--(j) show the  initial part of the forecast and correspond to (a)--(e).}
	\label{fig:results_timecoeff}
\end{figure*}
%------------------------------------------------------
%------------------------------------------------------
\begin{figure}[!hptb]
	\includegraphics[width = 8.5cm]{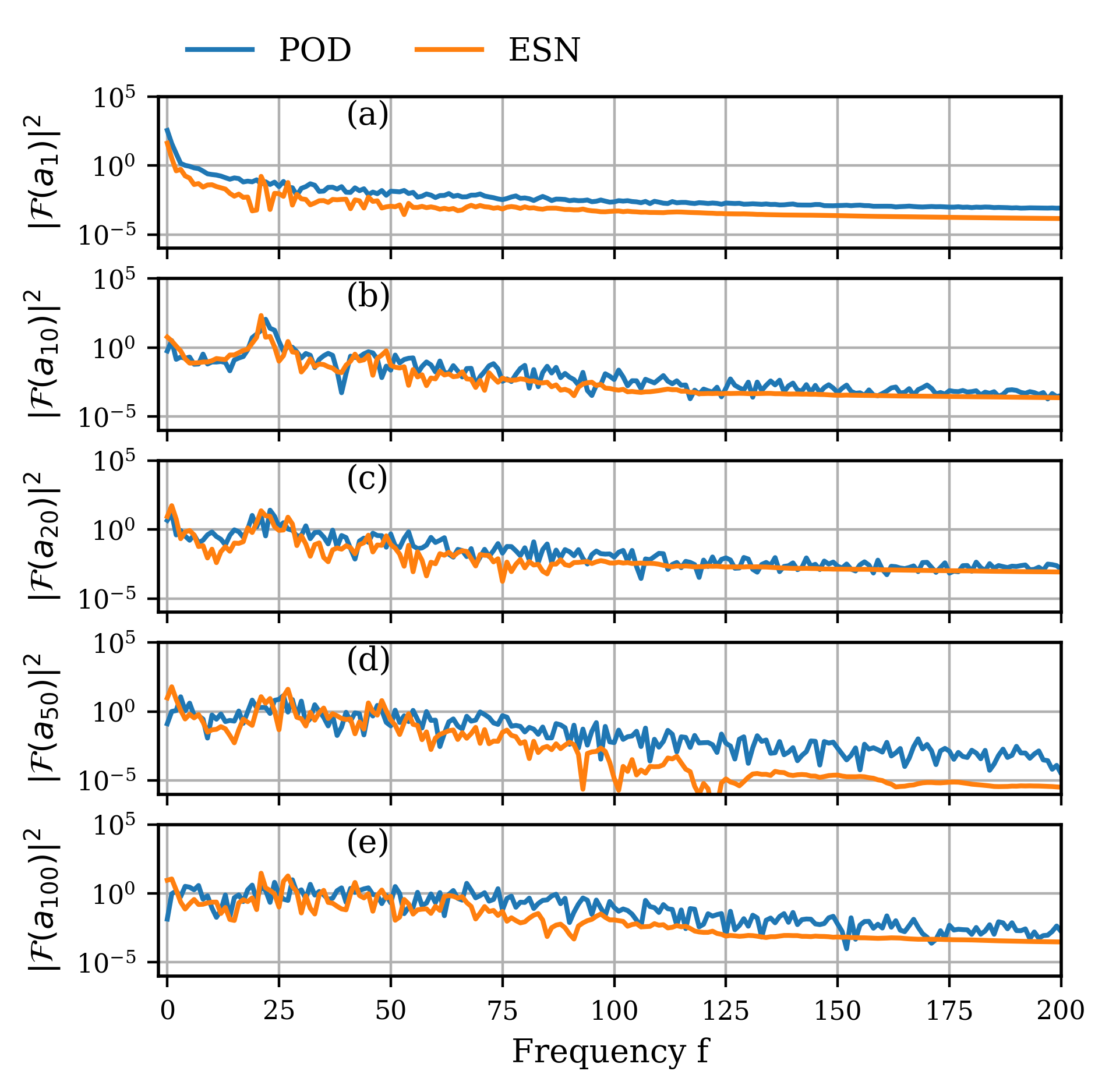}
	\caption{Fourier spectrum $\|\mathcal{F}(a_i)\|^2$ of the POD time coefficients and of the corresponding reservoir prediction. The first 200 frequencies are shown only.}
	\label{fig:results_timecoeff_fft}
\end{figure}
%------------------------------------------------------
After the ESN receives the initial input $\mathbf{u}(0)=\mathbf{a}(0)$, the autonomous predictor (see eq. (\ref{eq:rcm_couple})) produces estimates for the POD time coefficients which can be seen in Fig. \ref{fig:results_timecoeff}.  From the predicted coefficients and the known POD modes we can reconstruct all fields and compare these with the ground truth which is the POD expansion of the test data.

Deviations for the first ten coefficients are detected while predictions for subsequent POD coefficients associated with less energetic modes agree with the values of the validation data for the first few time steps. Nevertheless, the ESN accomplishes to produce a time series with matching temporal frequency as the actual data, but shows bigger deviations to compute the trend of the slowly varying first coefficients.  

The frequency spectra of the ESN predictions for the coefficients $a_i(t)$ in comparison to those of the test data are shown in Figure \ref{fig:results_timecoeff_fft} for 5 different cases. While the spectral values of the first 100 frequencies are captured well by the ESN, the higher frequency part starts to deviate in most cases. As discussed already above, this might be due to a simple RC model architecture, which does not differentiate between significantly different time scales that are always present in a turbulent flow. Nevertheless, the result underlines that the ESN is able to learn the time scales of the most significant POD coefficients.

\begin{figure*}[!hptb]
	\includegraphics[width = 14cm]{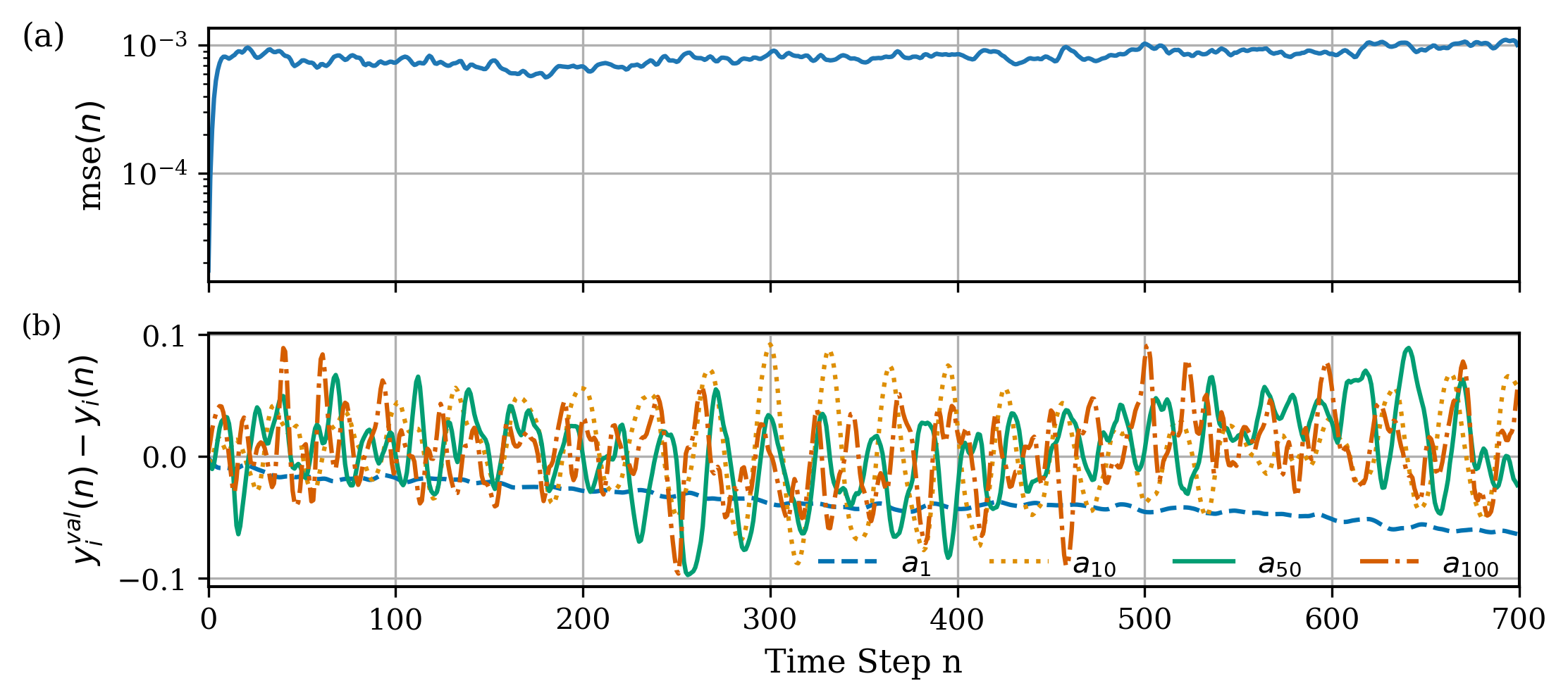}
	\caption{ESN prediction error: (a) Mean square error over all modes ${\rm mse}(n)$, see. eq. (\ref{eq:mode_mse}), a function of time steps $n$ after the training phase.  (b) The difference $(y_i^{\rm val}(n)- y_i(n))$, i.e. the deviation of the ESN prediction $y_i(n)$ of $a_i$ at time steps $n$ after training.}
	\label{fig:results_error}
\end{figure*}

Figure \ref{fig:results_error}(a) shows the mean square error ${\rm mse}(n)$ over all modes, as defined in (\ref{eq:mode_mse}), as a function of time steps after training. The mean error initially rises and then saturates. The fact that errors increase stems from the coupling scheme of output to input; small errors will inadvertently be amplified by the nonlinearity of the activation function in \eqref{eq:rcm_activation}. Figure \ref{fig:results_error}(b) shows the deviations $(y_i^{\rm val}(n)- y_i(n))$  for $i = 1,10,50,100$.
%------------------------------------------------------
%------------------------------------------------------

\begin{figure*}[!hptb]
	\includegraphics[width = 16cm]{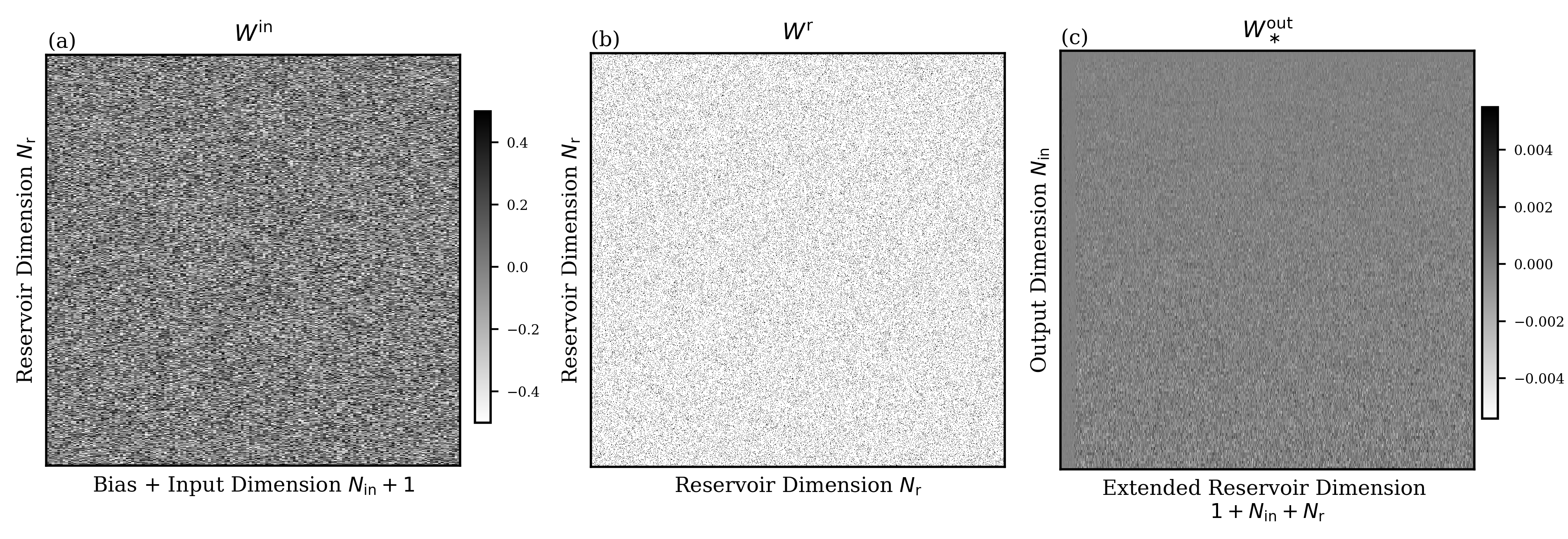}
	\caption{Reservoir weight matrices: (a) Input weight matrix $W^{\rm in}$ which is a $4000\times 151$ matrix in our case. (b) Reservoir weight matrix $W^{\rm r}$ which is a $4000\times 4000$ matrix in the present case. All weights that are unequal to zero are marked as black dots. (c) Optimized output weight matrix $W_\ast^{\rm out}$ which is a $150\times 4151$ matrix. The aspect ratios of $W^{\rm in}$ and $W_\ast^{\rm out}$ have been adjusted for illustration purposes.}
	\label{fig:results_weights}
\end{figure*}

Figures \ref{fig:results_weights}(a)--(c) show the three weight matrices of the RC model. As described in section \ref{subsec:rc}, the input and reservoir weights are initialized randomly and left unchanged. With a reservoir density of $D=0.2$ the reservoir matrix $W^{\rm r}$ is a sparse matrix containing many weights equal to zero. Note further, that the fitted output weights $W^{\rm out}$ have low magnitudes in comparison to the entries of the input matrix $W^{\rm in}$. This is adjusted according to the number of reservoir nodes $N_{\rm r}$ and the magnitude of the training data. Moreover, the magnitude of the first $1+N_{\rm in}$ columns are close to zero. This indicates that the contributions of the output bias $b$ and the current input $\mathbf{u}$ to the reservoir output (see eq. (\ref{eq:rcm_out})) are small. 

\begin{figure*}[!hptb]
	\includegraphics[width = 14cm]{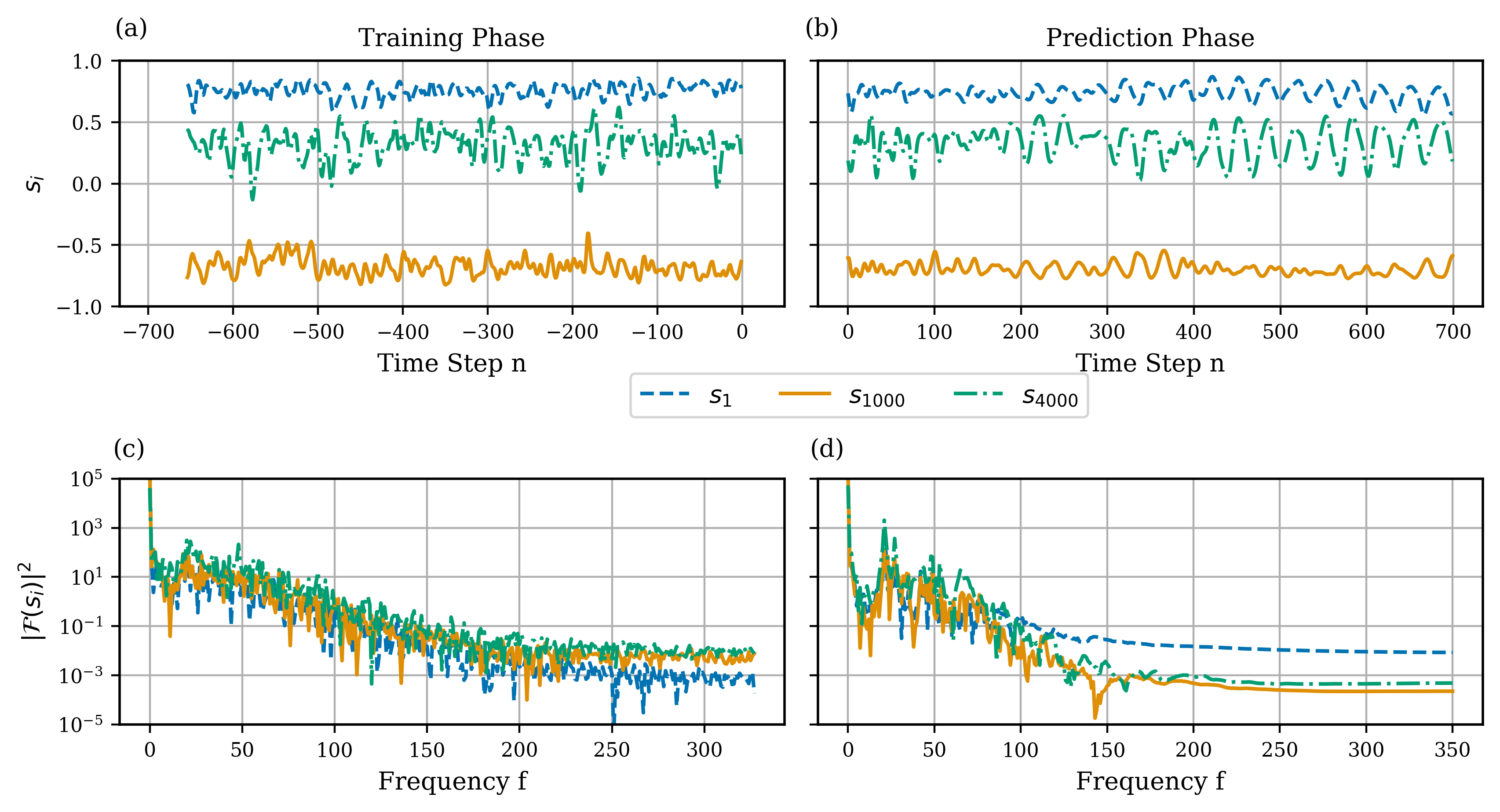}
	\caption{Reservoir state components $s_1,s_{1000}, s_{4000}$ versus time step $n$ during (a) training and (b) prediction phases. Panels (c) and (d) show their corresponding Fourier spectra $\|\mathcal{F}(s_i)\|^2$. Note that in (a), the first $46$ time steps  are not shown, as they were used for the initialization of the reservoir.}
	\label{fig:result_states}
\end{figure*}

Figure \ref{fig:result_states} (a,b) shows the training and prediction phase dynamics of three exemplary hidden reservoir state components $s_1, s_{1000}$, and  $s_{4000}$. As the first $46$ time steps of the training data were used to initialize the reservoir state, the last $n_{\rm Train}-46$ time steps are shown in panel (a) only. During both phases the individual time series $s_i$ are confined to a certain subrange of the whole range $\left[-1,1\right]$. They have comparable amplitudes. Nevertheless, the prediction phase time series differ from the their training phase counterparts by slightly smoother variations with respect to time, see also the corresponding Fourier frequency spectra in panels (c,d) of the same figure. This might be explained by the fact that the states for $n\ge 0$ experience the learned output matrix $W^{\rm out}_{\ast}$ via the closed feedback loop. The states in the training phase, $n<0$, on the other hand, do neither experience the fitted output matrix, nor is the last output fed back to the reservoir. We suspect that this has a significant impact on the evolution of the $s_i$.
%------------------------------------------------------
%------------------------------------------------------
%------------------------------------------------------

\begin{figure*}[!hptb]
	\includegraphics[width = 440pt]{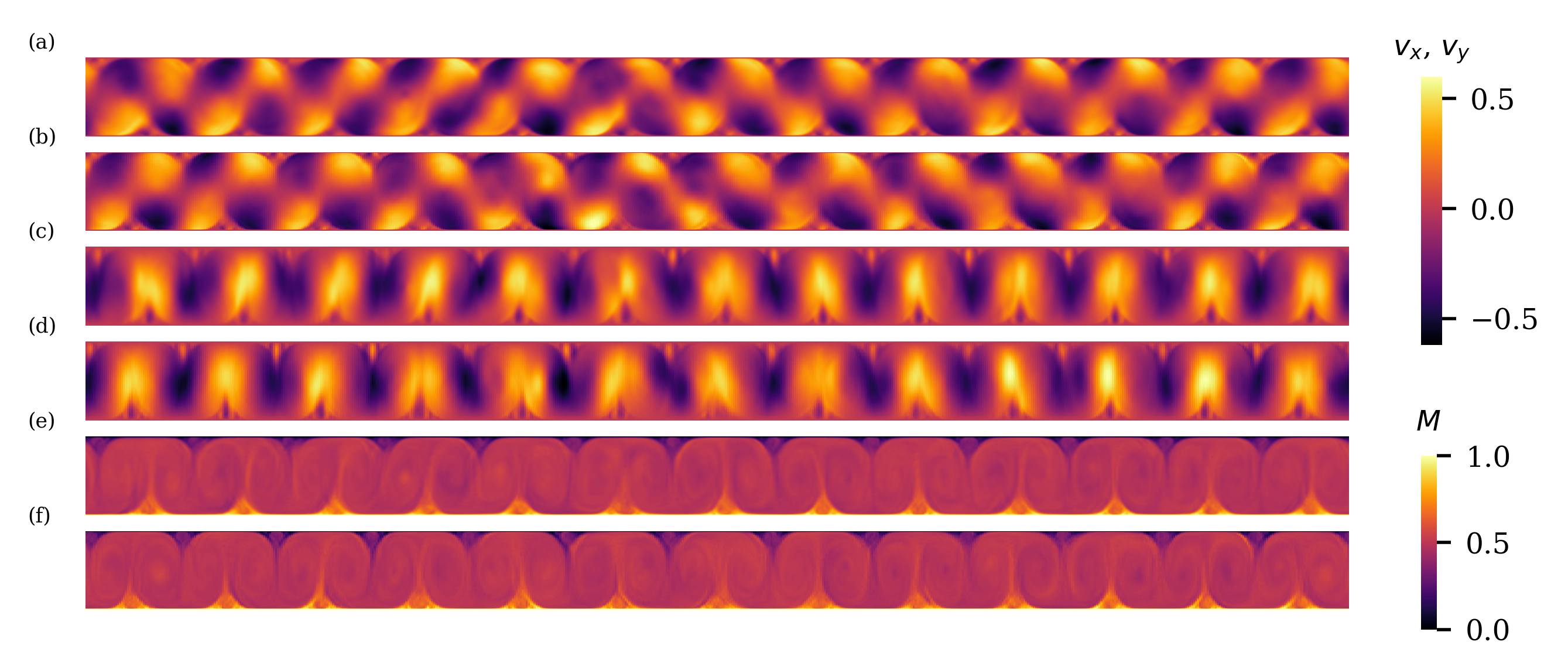}
	\caption{Instantaneous snapshot of the fields (a, b) $v_x(x,y)$, (c, d) $v_y(x,y)$ and (e, f) $M(x,y)$ at time step $n=350$ after the training phase. Panels (a, c, e) are validation data from the POD and (b, d, f) the ESN output data. The fields were reconstructed using the first 150 $a_{p}(n)$ (POD) and the predictions $y_p(n)$ (ESN). Here, $n$ is a discrete time step. The ESN predictions deviate locally from the POD fields, but capture large-scale features of the flow. The aspect ratio has been adjusted again for illustration purposes. The corresponding colorbars can be seen on the right.}
	\label{fig:results_fields}
\end{figure*}

We now take a look at the reconstruction of the three fields $v_x(x,y)$, $v_y(x,y)$ and $M(x,y)$ with the reservoir outputs as temporal coefficients to see whether large-scale features are captured correctly. An instantaneous snapshot at the half-time of the prediction phase at time step $n=350$ is depicted in Fig. \ref{fig:results_fields}. We apply a POD-mode  composition (\ref{eq:pod}) to obtain the fluctuation fields $v_x'$, $v_y'$ and $M'$ from the reservoir outputs $a_p(t)$. The mean profiles $\langle v_x \rangle_t$ and $\langle v_y \rangle_t$ and $\langle M\rangle_t$ are subsequently added to obtain the full fields, see eqns. (\ref{eq:reynolds_ux}) and (\ref{eq:reynolds_M}). The resulting ESN predictions are displayed in panels (b), (d), and (f). For reference, the validation (POD) fields are shown in panels (a), (c), and (e). 

The horizontal velocity field $v_x(x,y)$ in (a) and (b) shows some differences in the structure of the right- and left-going fluid patches, but the large-scale structure as a whole is in surprisingly good qualitative agreement. 
The structure of vertical velocity field $v_y(x,y)$ in panel (c) and (d) does not show a systematic distinction, even though slight differences in shape and maximum values of up- and downdrafts are detectable. Finally the moist buoyancy field $M(x,y)$ in (f) does not fully reproduce all moist plumes that detach from the bottom plate, see validation field in (e). Nevertheless the predicted time coefficients lead to reconstructed fields that contain the same features as the original fields. 
%------------------------------------------------------
%------------------------------------------------------

\begin{figure}[!hptb]
	\includegraphics[width = 200pt]{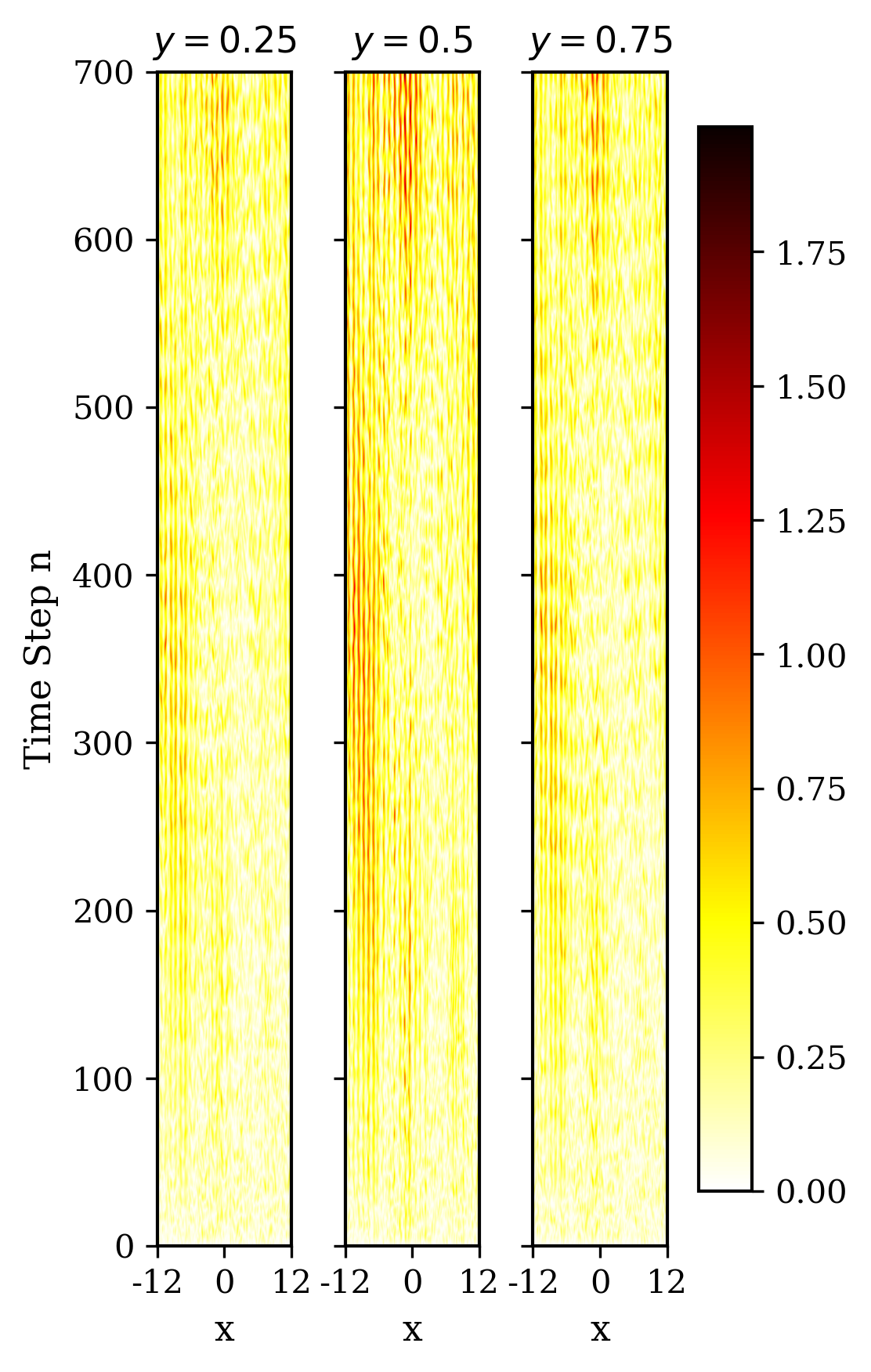}
	\caption{Time evolution of the prediction error ${\rm Err}(x,n)$, which is given by eq. \eqref{error}, at specified height $y$ (see top). As time progresses, the errors start to grow in magnitude. Different positions $(x,y)$ in the domain give rise to different magnitudes of the deviation.}
	\label{fig:results_errorpropagation}
\end{figure}

To get a better grasp on the time evolution of the error in the field reconstruction, we compute a normalized field deviation at constant height $y$ of the vertical velocity field component in Fig. \ref{fig:results_errorpropagation} which is given by
\begin{align}
{\rm Err}(x,n)=\frac{| v_y'^{\rm POD} (x,n)-v_y'^{\rm ESN}(x,n)|}{\max_{x,y,n}\left(v_y'^{\rm POD}\right)}\Bigg|_{y={\rm const}}
\label{error}
\end{align}
where the superscript defines whether the field was reconstructed with $a_i(n)$ (POD) or $y_i(n))$ (ESN).  We find that the fast growing errors in the time coefficients lead to fast amplifications of local field errors. Furthermore, different horizontal and vertical positions in the domain show different error magnitudes.
%------------------------------------------------------
%------------------------------------------------------

\begin{figure}[!hptb]
	\includegraphics[width = 160pt]{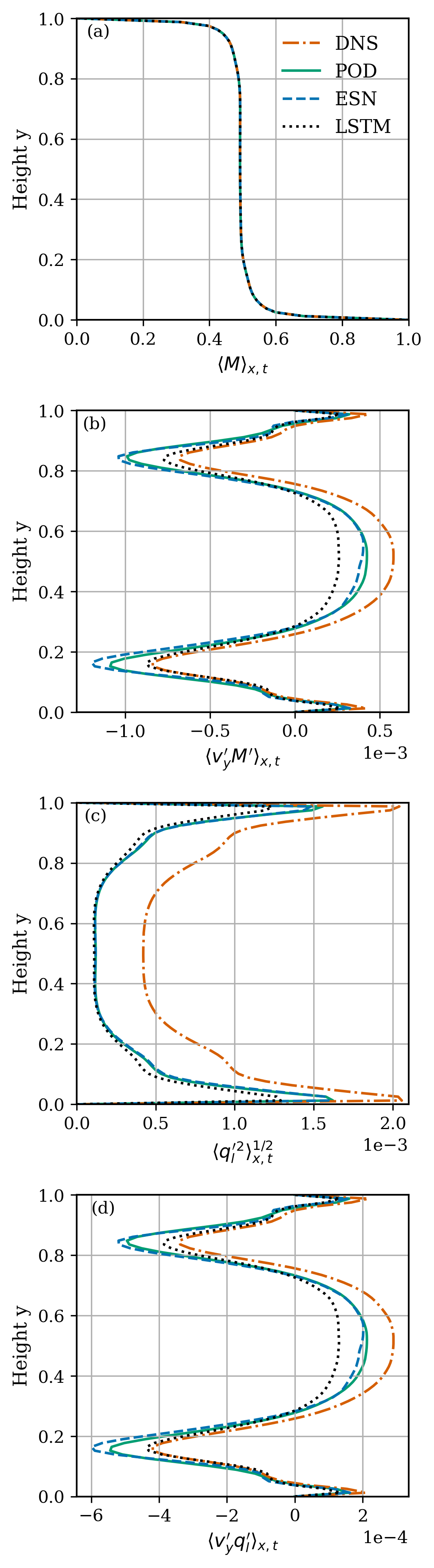}
	\caption{Line-time averaged vertical profiles $\langle \cdot \rangle_{x,t}$ of (a) the full moist buoyancy $M$, (b) the vertical moist buoyancy flux $v_y'M'$, (c) the fluctuations of the liquid water content $q_l$ and (d) the vertical liquid water flux $v_y'q_l'$. The time average for DNS (dash-dotted) and POD (solid) were computed over the whole range of 1400 time steps, while for the ESN (dashed) and LSTM (dotted) only the range of the prediction phase ($700$ time steps) was taken into account.}
	\label{fig:results_verticalprofiles}
\end{figure}

We now discuss the ability of the ESN to reproduce the low-order statistical properties of the turbulent flow. This is done by comparison of vertical line-time average profiles $\langle \cdot \rangle_{x,t}(y)$. The averages are taken along $x$-direction in combination with respect to time $t$. Such profiles are for example of interest in larger-scale atmospheric models for the parameterization of sub-grid scale transport~\cite{grabowski_coupling_2001,khairoutdinov_cloud_2001}. Figure \ref{fig:results_verticalprofiles} depicts the profiles as a function of the domain height $y$. The actual profiles obtained by the original DNS are plotted as a dash-dotted,  the POD reconstruction as a solid and the reconstruction from the ESN outputs as a dashed line. Figure \ref{fig:results_verticalprofiles}(a) shows the moist buoyancy $M$. Here the time mean $\langle M\rangle_{t}$ was added to see whether the reconstructed POD and ESN fields would deviate from the full DNS data. We observe an excellent agreement and find that the ESN produces correct fluctuations which preserve this profile. The fluctuations of the vertical moist buoyancy flux $\langle v_y'M'\rangle$ are shown in Fig. \ref{fig:results_verticalprofiles}(b). Again, an excellent agreement between the curves in the bulk of the domain and small deviations in the boundary layers only are observable, despite the fact that this quantity is much more susceptible to errors since it consists of two ESN estimates. Finally the fluctuations from the liquid water content and the liquid water flux, further derived fields, are shown in \ref{fig:results_verticalprofiles}(c) and (d), respectively. Here we see that POD and ESN curves match throughout the whole of the domain. We thus conclude that the ESN is able to reproduce essential low-order statistics very well. 

For comparison, we add a comparison of test data with the output of an LSTM network. The network parameters are as follows: the number of hidden states is 300 and the number of stacked LSTM layers 3. The loss function is again the mean-squared error, the optimization applies the method of adaptive moments, and the learning rate is $10^{-3}$ \cite{Goodfellow-et-al-2016}.  A training of the network proceeds over 1000 epochs. We can conclude that the LSTM performs similarly well as the ESN even though the reproduced profiles deviate a bit for both fluxes in the center of the convection layer.
%------------------------------------------------------
%------------------------------------------------------
%------------------------------------------------------
\begin{table*}[!hptb]
	\begin{ruledtabular}
		\begin{tabular}{c c c c c c}
			$\langle {\rm CC}^{\rm DNS}\rangle_t$ & $\langle q_l^{\rm DNS}\ge0\rangle_{x,y,t}\cdot 10^{3}$ &$\langle {\rm CC}^{\rm POD}\rangle_t$ &  $\langle q_l^{\rm POD}\ge0\rangle_{x,y,t}\cdot 10^{3}$ & $\langle {\rm CC}^{\rm ESN}\rangle_t$& $\langle q_l^{\rm ESN}\ge0\rangle_{x,y,t}\cdot 10^{3}$\\
			\hline
			$89.43$\% &$3.24$ & $83.25$\% &$2.72$ & $82.49$\%&   $2.72$\\
		\end{tabular}
	\end{ruledtabular}
	\caption{Time mean average $\langle \rm CC\rangle_t$ and $\langle q_l\ge 0\rangle_{x,y,t}$ of the cloud cover CC and the volume average of liquid water for the DNS, POD and ESN case. See also Fig. \ref{fig:results_clouds}.}
	\label{tab:cloud_cover}
\end{table*}
%------------------------------------------------------
%------------------------------------------------------
\begin{figure*}[!hptb]
	\includegraphics[width = 440pt]{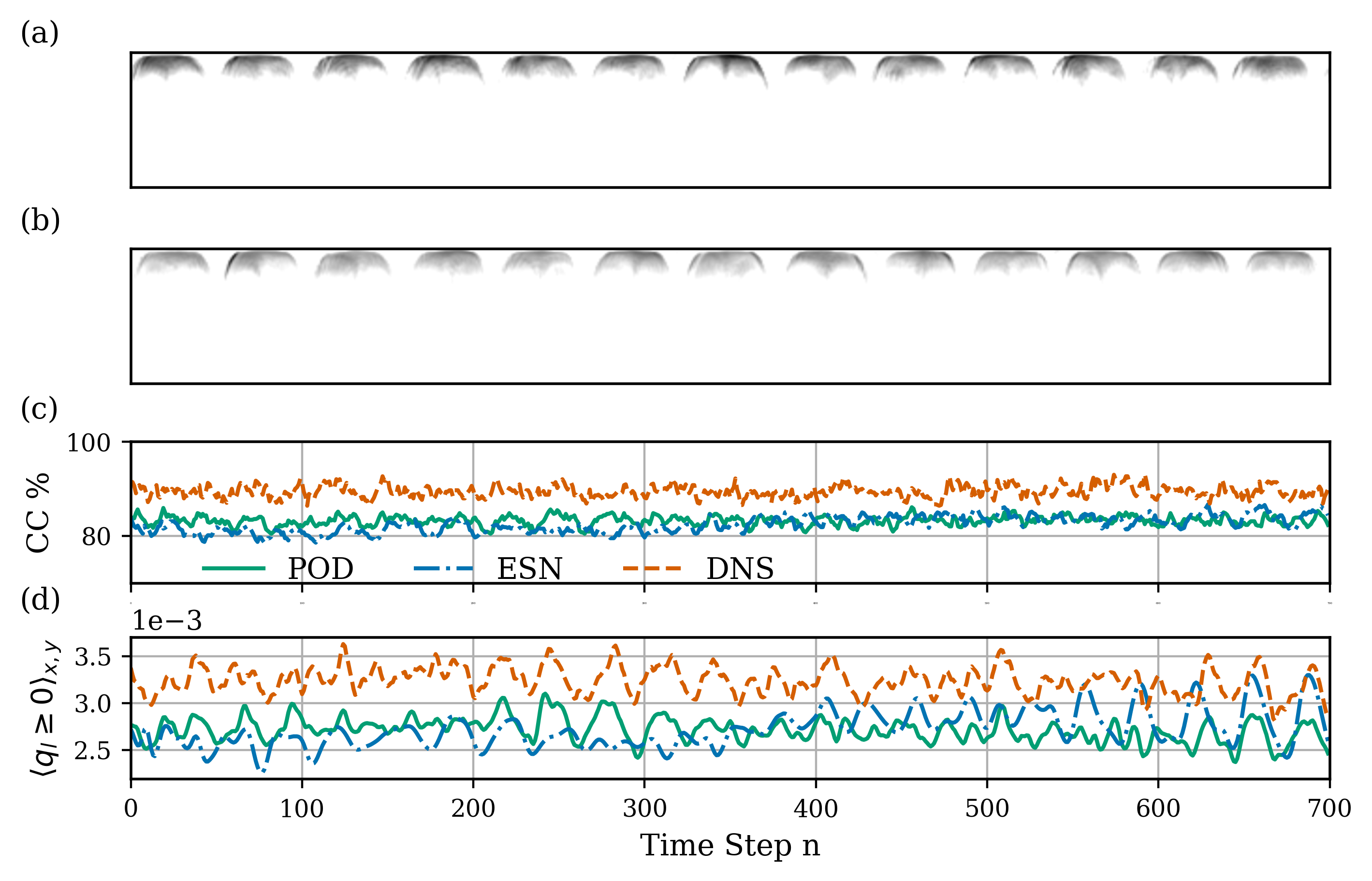}
	\caption{Clouds, i.e. $q_l(x,y,n) \ge 0$ at time step $n=350$ of (a) the POD and the (b) ESN prediction. The liquid water content $q_l$ differs in the magnitude and shape of its envelope, the isosurface $q_l=0$. (c) shows the cloud cover as defined in eq. (\ref{eq:cloud_cover}) computed for the DNS data (brown), POD data (green) and the ESN predictions (blue). The POD approximation does not capture all of the original cloud cover; the value of the DNS exceeds the one of the POD by a few per cent. The cloud cover prediction of the ESN itself deviates by a few per cent in comparison to that of the POD. (d) shows the change of the volume average of positive liquid water content $\langle q_l\ge 0\rangle_{x,y}$ with time.}
	\label{fig:results_clouds}
\end{figure*}
%------------------------------------------------------

Motivated by these results, we now investigate whether quantities such as the cloud cover can be also modeled by the ESN. We define the cloud cover CC of the two-dimensional domain $N_x'\times N_y'$ as the ratio of the number of vertical grid lines $N_{q_l>0}$ that contain at least one mesh point with $q_l>0$ along their vertical line of sight and the total number of vertical grid lines, $N_x'$. Thus follows 
\begin{equation}
{\rm CC}  = \frac{N_{q_l>0}}{N_x'}\times 100 \%.
\label{eq:cloud_cover}
\end{equation}
The time average $\langle {\rm CC}\rangle_t$ and volume-time average of positive liquid water content $\langle q_l\ge 0\rangle_{x,y,t}$ are given in Table \ref{tab:cloud_cover}. The truncation to the first $\rm N_{\rm POD}$ POD modes leads to a loss of about $6.9\%$ of the original DNS CC and a $16.1\%$ loss of $\langle q_l\ge 0\rangle_{x,y,t}$. We find good agreement between ESN estimate and POD results. In Fig. \ref{fig:results_clouds}, the POD and ESN results of the cloud distribution at time step $n=350$ are shown. Despite small discrepancies in the local distribution of the liquid water content and the shape of the cloud boundaries, i.e. the isosurfaces $q_l=0$, the overall distribution is comparable. In panels (c) and (d) of the same figure, the time evolution of cloud cover and volume-time average positive liquid water content are displayed. While the predicted cloud cover does not deviate too much from the reference case, the variations in the amount of liquid water are less well reproduced. 
 
\section{Summary and Conclusion}
\label{sec:conclusion}
In the present work, we have applied a machine learning algorithm to two-dimensional turbulent moist Rayleigh-Bénard convection, in order to infer the large-scale evolution and low-order statistics of convective flow undergoing phase changes. We apply a specific learning scheme for recurrent neural networks, called echo state network, which has been applied for  learning the dynamics of nonlinear systems, such as turbulent shear flows. Here, we test its capabilities successfully by fitting a reservoir to complex convection flow dynamics which results by the interaction of turbulence with the nonlinear thermodynamics originating from the first order phase changes between vapor and liquid water as present for example in atmospheric clouds. We therefore generate comprehensive data by means of a simple moist convection model in the Boussinesq framework, the moist Rayleigh-B\'{e}nard convection model. 

We obtain moist convection data from direct numerical simulations. As the 2d set of data has still a large amount of degrees of freedom and therefore cannot be passed directly to the echo state network, we introduce the POD as an additional dimensionality reduction step. We therefore decompose the data into a data-driven, spatially dependent basis with temporal coefficients, where the latter are fed to the reservoir. We truncate the POD to its most energetic modes and coefficients, reducing the degrees of freedom of the dynamical system at hand considerably. This reduced data set serves as the ground truth for the training of the echo state network as well as validation of its outputs. The network setup is tuned by conducting a comprehensive grid search of important hyperparameters. By coupling the output of the trained network back to its input, the autonomous system estimates the evolution of the temporal coefficients. Reconstructing the velocity and thermodynamic fields from these estimates, allow us to check whether the dynamics have been learned. We find an excellent agreement of the vertical profiles of moist buoyancy, vertical moist buoyancy transport as well as liquid water content. Furthermore, we report a good agreement of essential parameters in moist convection such as the fraction of clouds covering the two-dimensional atmosphere, as well its content of liquid water. 

This first approach of our reservoir computing model to moist Rayleigh-Bénard convection shows, its potential to infer low-order statistics from a set of training data. Though the reservoir output quickly diverges from the actual system trajectory, time averaged quantities are robustly reproduced. This result might seem trivial at first glance, yet the reservoir produces velocity and thermodynamic fluctuation fields which do not deviate too strongly from those of the original flow, even for combined quantities such as the liquid water flux across the layer. This indicates that the present echo state network did not just learn the statistics, but the dynamical system itself. Our additional comparison with an LSTM network gives a similar outcome. A more detailed comparison of both RNN implementations has to be left however as a future work.

Our approach can be considered as a first step of applying reservoir computing as a machine learning-based parameterization. General circulation models already use multi-scale modeling methods where small-scale resolving models interact with the large-scale motion by their low-order statistics, essentially relaxing one to each other, e.g. in superparametrizations~\cite{grabowski_crcp_1999,grabowski_coupling_2001, khairoutdinov_cloud_2001}. An echo state network can serve as a simple dynamical substitute for the unresolved subgrid scale transport.

Even though the present results are promising, the development is still in its infancy. We state that for example the mathematical foundations of reservoir computing, which could provide deeper insights on the role of the hyperparameters on the prediction quality, are still mostly unexplored. Moreover, we reckon that for an extension of the ESN approach to a three dimensional flow, the data reduction step via POD will not suffice to cope with the large amount of simulation data. For this scenario one might propose the usage of a convolutional autoencoder/-decoder neural network in combination with the RC model. Furthermore, we mention that the machine learning algorithm is supposed here to learn dynamics of a nonlinear system which incorporates processes on different spatial and temporal scales. This circumstance is so far not fully captured by the network architecture. Particularly for turbulence, this might imply that the different spatial scales which interact with each other and exchange their energy, could be trained separately allowing for a subsequent coupling. The exploration of such ideas is currently under way and will be reported elsewhere.

\vspace{0.3cm}
\section*{Acknowledgments}
This work is supported by the project "DeepTurb -- Deep Learning in and of Turbulence" which is funded by the Carl Zeiss Foundation. The authors gratefully acknowledge the Gauss Centre for Supercomputing e.V. (\url{www.gauss-centre.eu}) for funding this project by providing computing time through the John von Neumann Institute for Computing (NIC) on the GCS Supercomputer JUWELS at J\"ulich Supercomputing Centre (JSC). We thank Martina Hentschel, Erich Runge, and Sandeep Pandey for helpful comments.

%% ------------------------------------------------------------------------ %%
%% References and Citations

%%%%%%%%%%%%%%%%%%%%%%%%%%%%%%%%%%%%%%%%%%%%%%%
%%%%%%%%%%%%%%%%%%%%% BIBLIOGRAPHY%%%%%%%%%%%%%%%%%%
%%%%%%%%%%%%%%%%%%%%%%%%%%%%%%%%%%%%%%%%%%%%%%%%

\bibliography{HeySchu_rev2}

\end{document}